\documentclass[iop,superscriptaddress]{emulateapj}

\usepackage{graphicx}
\usepackage{dcolumn}
\usepackage{amssymb} 
\usepackage{ctable}
\usepackage{bm}
\usepackage{epsfig}
\usepackage{epstopdf}
\usepackage{epsf,color}
\usepackage{natbib}

\newcommand{\cmnt}[1]{}

\def\VEV#1{\left\langle #1 \right\rangle}

\newcommand{\beqa}{\begin{eqnarray}}
\newcommand{\eeqa}{\end{eqnarray}}

\begin{document}

\title{Cross-correlations as a Cosmological Carbon Monoxide Detector}

\author{Anthony R. Pullen}
\affiliation{NASA Jet Propulsion Laboratory, California Institute of Technology, 4800 Oak Grove Drive, MS 169-237, Pasadena, CA, 91109, U.S.A.}
\author{Tzu-Ching Chang}
\affiliation{IAA, Academia Sinica, P.O. Box 23-141, Taipei 10617, Taiwan}
\author{Olivier Dor\'{e}}
\affiliation{NASA Jet Propulsion Laboratory, California Institute of Technology, 4800 Oak Grove Drive, MS 169-215, Pasadena, CA, 91109, U.S.A.}
\affiliation{California Institute of Technology, MC 249-17, Pasadena, California, 91125 U.S.A.}
\author{Adam Lidz}
\affiliation{Department of Physics and Astronomy, University of Pennsylvania, 209 South 33rd Street, Philadelphia, PA, 19104, U.S.A.}

\email{anthony.r.pullen@jpl.nasa.gov}

\begin{abstract}
We present a new procedure to measure the large-scale carbon monoxide (CO) emissions across cosmic history. As a tracer of large-scale structure (LSS), the CO gas content as a function of redshift can be quantified by its three-dimensional fluctuation power spectra. Furthermore, cross-correlating CO emission with other LSS tracers offers a way to measure the emission as a function of scale and redshift.  Here we introduce the model relevant for such a cross-correlation measurement between CO and other LSS tracers, and between different CO rotational lines.  We propose a novel use of Cosmic Microwave Background (CMB) data and attempt to extract redshifted CO emissions embedded in the Wilkinson Microwave Anisotropy Probe (WMAP) dataset. We cross-correlate the all-sky WMAP7 data with LSS data sets, namely, the photometric quasar sample and the luminous red galaxy sample from the Sloan Digital Sky Survey Data Release 6 and 7 respectively.  We are unable to detect a cross-correlation signal with either CO(1-0) nor CO(2-1) lines, mainly due to the instrumental noise in the WMAP data.  However, we are able to rule out models more than three times greater than our more optimistic model. We discuss the cross-correlation signal from the thermal Sunyaev-Zeldovich effect and dust as potential contaminants, and quantify their impact for our CO measurements. We discuss forecasts for current CMB experiments and a hypothetical future CO focused experiment, and propose to cross-correlate CO temperature data with the Hobby-Eberly Telescope Dark Energy Experiment Ly$\alpha$-emitter sample, for which a signal-to-noise ratio of 58 is possible.

\end{abstract}
\maketitle

\section{Introduction} \label{S:intro}

The correlation of redshifted emission line spectra has been proposed as a potentially powerful probe of large-scale structure (LSS) at high redshifts \citep{1999ApJ...512..547S,2008A&A...489..489R,2010JCAP...11..016V}.  While traditional LSS probes such as galaxies and quasars for the foreseeable future will be limited to $z<2$ and $z<4$, respectively, mapping emission lines can probe out to $z>6$, potentially mapping the Epoch of Reionization (EoR).  In these very high-redshift regions, emission lines can also be cross-correlated with the 21-cm line from the intergalactic medium (IGM), constituting a very complementary probe to 21-cm alone or extragalactic emission lines alone, whose auto-spectra are plagued with foregrounds \citep{2011ApJ...728L..46G,2011ApJ...741...70L}.  Three extragalactic emission lines of interest in these types of studies include CO \citep{2008A&A...489..489R,2011ApJ...730L..30C,2011ApJ...728L..46G,2011ApJ...741...70L}, CII \citep{2012ApJ...745...49G}, and HI \citep{2011ApJ...728L..46G}.  In this paper, we will focus on the CO lines.

CO is readily produced from carbon and oxygen in star-forming regions.  The CO molecule exhibits several rotational transitional states $J\to J-1$ with line frequencies $\nu_J=J\nu_{\rm CO}$ and $\nu_{\rm CO}=115$ GHz.  CO lines have been studied both as foreground contaminants to Cosmic Microwave Background (CMB) observations \citep{2008A&A...489..489R} as well as LSS tomography probes \citep{2011ApJ...728L..46G,2011ApJ...741...70L}.  In this paper we will follow the formalism presented in \citet{2011ApJ...741...70L} (hereafter L11)  and extend it to lower redshifts, producing two models that make different assumptions. The two CO rotational lines we seek to constrain are CO(1-0) and CO(2-1) because they are typically the brightest and both their emitting and redshifted frequencies conveniently fall in the frequency range of CMB observations, where vast amounts of data and expertise have been cultivated over the years.  The multiple emission lines of CO due to multiple rotational states allow interlopers to be identified more easily, giving CO an advantage over other emission lines as cosmological probes.  These observational prospects have created considerable interest in building future dedicated experiments, even if the exact strength of the signal at high redshift and on large scales is still rather uncertain. As such, we investigate what can be probed with current experiments.  For the LSS study of interest, we work in the regime of ``intensity mapping'', measuring aggregated CO emissions associated with the large-scale structure rather than with individual galaxies, as an efficient way of extracting the faint signals (e.g., \citet{2010Natur.466..463C}).  This approach is complementary to high-resolution CO observations enabled by the Atacama Large Millimeter/submillimeter Array (ALMA\footnote{\texttt{https://almascience.nrao.edu/}}) and the Jansky Very Large Array (JVLA\footnote{\texttt{https://safe.nrao.edu/evla/nova/index.shtml}}), for example, where individual galaxies have been mapped to high redshifts; their small (sub-arcminute) field-of-view is not well suited for large scale surveys.

One probe that could possibly constrain CO line emission is in fact the Wilkinson Microwave Anisotropy probe (WMAP) \citep{2011ApJS..192...14J}.  With band frequencies ranging from 23-94 GHz, WMAP should contain emission up to $z\sim 4$ from CO(1-0) and up to $z\sim 7$ from CO(2-1).  Constructing an auto-correlation spectrum in each band is very difficult due to contamination from other astrophysical processes that motivated WMAP in the first place such as the Cosmic Microwave Background (CMB) or galactic emission.  However, a cross-correlation with a different LSS tracer would not have the same foregrounds and thus be immune to these contaminants.  A suitable choice would be to cross-correlate the WMAP bands with photometric quasars in its corresponding CO redshift range.  Specifically, we choose the photometric quasar sample constructed from the Sloan Digital Sky Survey (SDSS) \citep{2000AJ....120.1579Y} Data Release 6 (DR6) \citep{2008ApJS..175..297A} in \citet{2009ApJS..180...67R}.  This sample contains quasars with redshifts as high as $z=6.1$, which covers most of WMAP's CO redshift range, allowing us to perform this cross-correlation for both CO(1-0) and CO(2-1) lines.  Note that this redshift range could allow us to search for higher CO lines.  CO(3-2) would be seen in V and W bands, while CO(4-3) and CO(5-4) would be seen only in the W band.   We leave searches for these lines to future work as they are expected to be dimmer. At redshifts $z<1$, we can cross-correlate the WMAP W band with the SDSS Data Release 7 (DR7) \citep{2009ApJS..182..543A} spectroscopic Luminous Red Galaxies (LRGs) sample in \citet{2010ApJ...710.1444K}.

In this paper, we propose cross-correlating brightness temperature maps with quasars (QSOs) and Luminous Red Galaxies (LRGs) to detect fluctuations in CO at high redshifts.  We first see what can be done with current data by cross-correlating WMAP temperature maps in each band with photo-quasars from SDSS DR6 and LRGs from SDSS DR7 in the band's CO redshift range.  We attempt to remove CMB fluctuations from the maps to isolate the CO signal (see Sec.~\ref{S:method} for details).  We report no detection of a cross-correlation between quasars or LRGs and CO temperature due to extragalactic CO line emission, although we place an upper limit that rules out any brightness level much greater than our model.   It appears that the weakness of these constraints is mainly due to instrumental noise in the WMAP temperature maps, although the density of quasars plays a lesser role.  We do not suspect quasar photo-z errors to play a role since the errors are $\Delta z\sim O(0.1)$ while the redshift bins we use are of order unity. To detect these lines in the future, we look towards a hypothetical future CO experiment to measure CO fluctuations for the line CO(1-0) at $z=3$ over the range $2.9<z<3.1$ with 20 redshift bins.  We forecast that this experiment cross-correlated with the full spectroscopic quasar survey from the Baryon Oscillation Spectroscopic Survey (BOSS) \citep{2011AJ....142...72E,2012ApJS..199....3R} will measure the cross-correlation amplitudes 
with signal-to-noise ratios (SNR) as high as 4 for each redshift bin and 20 for the entire redshift range, depending on the model.  The CO experiment cross-correlated with the {\it Hobby-Eberly Telescope Dark Energy Experiment} (HETDEX) \citep{2008ASPC..399..115H} could do significantly better with an SNR of 13 for each redshift bin and 58 for the entire redshift range.  We also set constraints for current and future ground-based CMB polarimeters like SPTPol or ACTPol.

The plan of our paper is as follows: in Sec.~\ref{S:data}, we describe the data products we use, including the WMAP 7-year observations (WMAP7) \citep{2011ApJS..192...14J} temperature maps, the photo-quasar maps from SDSS DR6, and the LRG maps from SDSS DR7.  In Sec.~\ref{S:eqns}, we derive the form of the CO-LSS cross-power spectra along with statistical errors for both models.  In Sec.~\ref{S:method}, we describe the estimator for the cross-correlation, presenting measurements in Sec.~\ref{S:results}.  Finally, we present forecasts for a possible future CO experiment as well as ground-based experiments in Sec.~\ref{S:forecasts} and conclude in Sec.~\ref{S:conclude}.  Wherever not explicitly mentioned, we assume a flat $\Lambda$CDM cosmology with parameters compatible with WMAP7.

\section{Data} \label{S:data}

\subsection{WMAP Temperature} \label{S:wmaptemp}

We attempt to extract redshifted CO emission associated with LSS embedded in the WMAP7 \citep{2011ApJS..192...14J}\footnote{http://lambda.gsfc.nasa.gov/product/map/current} temperature maps by means of cross-correlation.  Primordial CMB and Galactic foregrounds which contribute to the WMAP temperature maps will not correlate with LSS;  however, CMB secondary anisotropy signals, in particular the thermal Sunyaev-Zeldovich effect (tSZ) \citep{1980ARA&A..18..537S}, are expected to correlate with LSS. TSZ has been detected using similar data sets in \citet{2010ApJ...720..299C} and we will discuss it more in Sec.~\ref{S:results:SZ}.  Radio and dust emission also should correlate with LSS, and we discuss its contribution in Sec.~\ref{S:results:SZ} as well.  WMAP has 5 temperature bands: K (23 GHz), Ka (33 GHz), Q (41 GHz), V (61 GHz), and W (94 GHz) with bandwidths of 5.5, 7.0, 8.3, 14.0, and 20.5 GHz, respectively.  We do not use the K map since it is dominated by Galactic emission; thus we are left with 4 bands.  We use the HEALPix \citep{2005ApJ...622..759G} $N_{res}=9$ maps with the KQ85 mask to remove the brightest resolved point sources and the bright Galactic emission regions near the galactic plane. These cuts leave each map containing 2,462,208 $N_{res}=9$ pixels, covering 32,289 deg$^2$ (78.3\% of the sky).

One subtlety that must be taken into account is that the temperature values in the WMAP data products are perturbations of {\it physical} temperature, assuming a mean temperature equal to the CMB temperature.  The CO temperature, however, is a brightness temperature.  These two temperature measurements diverge at the higher frequency bands, mainly the V and W bands.  Therefore, before performing our analysis we must convert the CMB physical temperature perturbations to brightness temperature perturbations according to the formula in \citet{1997ApJ...480L...1G}
\begin{eqnarray}\label{E:brcon}
\delta T_b = \frac{x^2e^x}{(e^x-1)^2}\delta T
\end{eqnarray}
where $\delta T_b$ is the brightness temperature, $\delta T$ is the temperature map from the WMAP data product, and $x=h\nu/kT_{CMB}$ with $T_{CMB}=2.725K$ and $\nu$ being the band frequency.  The prefactor for $\delta T$ has a different value for each WMAP band, ranging from 0.97 for the Ka band to 0.80 for the W band. Note that this expression neglects the band width.

\subsection{SDSS Data}
We use photometric quasars from the SDSS DR6 \citep{2008ApJS..175..297A} and LRGs from the SDSS DR7 \citep{2009ApJS..182..543A} to trace the matter density and construct its angular power spectrum.  The SDSS consists of a 2.5-m telescope \citep{2006AJ....131.2332G} with a 5-filter (\textit{ugriz}) imaging camera \citep{1998AJ....116.3040G} and a spectrograph.  Automated pipelines are responsible for the astrometric solution \citep{2003AJ....125.1559P} and photometric calibration \citep{1996AJ....111.1748F, 2001AJ....122.2129H, 2006AN....327..821T, 2008ApJ...674.1217P}.  Bright galaxies, LRGs, and quasars are selected for follow-up spectroscopy \citep{2002AJ....124.1810S, 2001AJ....122.2267E, 2002AJ....123.2945R, 2003AJ....125.2276B}.  The data used here from DR6 and DR7 were acquired between August 1998 and July 2008.

\subsubsection{SDSS quasars}
We use the photometric quasar catalog composed by \citet{2009ApJS..180...67R} (hereafter RQCat).  The entire catalog consists of 1,172,157 objects from 8417 deg$^2$ on the sky selected as quasars from the SDSS DR6 photometric imaging data.  We limit our dataset in this analysis to UV-excess quasars (catalog column \textbf{uvxts} = 1) because they have a higher selection efficiency.  We also require the catalog column \textbf{good} $> 0$ to reject objects that are likely stars.  For the survey geometry we use the union of the survey runs retrieved from the SDSS DR6 CAS server.  We omitted runs 2189 and 2190 because many objects in these runs were cut from RQCat.  This mask was pixelized using the MANGLE software \citep{2004MNRAS.349..115H,2008MNRAS.387.1391S}.  We pixelize the quasars as a number overdensity, $\delta_q=(n_q-\overline{n})/\overline{n}$, onto a HEALPix pixelization of the sphere with $N_{res}=9$, where $n$ is the pixel's number of quasars divided by the pixel's survey coverage $w$ and $\overline{n}=(\sum_in_iw_i)/(\sum_iw_i)$.  We then reject pixels with extinction $E(B-V)\geq 0.05$, full widths at half-maximum of its point-spread function (PSF) FWHM $\geq 2$ arcsec, and stellar densities (smoothed with a $2^\circ$ FWHM Gaussian) $n_{stars}\geq 562$ stars/deg$^2$ (twice the average stellar density), similar to \citet{2008PhRvD..78d3519H}.  The extinction cut is very important because a high extinction affects the $u$ band, which is crucial to identifying quasars.  Also, since stars tend to be misidentified as quasars, it seems prudent to cut regions with high stellar density.  We implement these cuts using dust maps from \citet{1998ApJ...500..525S} and stars ($18.0<r<18.5$) from the SDSS DR6 \citep{2008ApJS..175..297A}.  We also reject pixels for which the survey region covers less than 80\% of the pixel area.   In addition, RQCat contains regions that seem undersampled.  We excise angular rectangles around these regions to remove them from the data.  The angular rectangles in celestial coordinates that were removed are $(\alpha,\delta)=(122^\circ\mbox{--}139^\circ,-1.5^\circ\mbox{--}(-0.5)^\circ)$, $(121^\circ\mbox{--}126^\circ,0^\circ\mbox{--}4^\circ)$, $(119^\circ\mbox{--}128^\circ,4^\circ\mbox{--}6^\circ)$, $(111^\circ\mbox{--}119^\circ,6^\circ\mbox{--}25^\circ)$, $(111.5^\circ\mbox{--}117.5^\circ,25^\circ\mbox{--}30^\circ)$, $(110^\circ\mbox{--}116^\circ,32^\circ\mbox{--}35^\circ)$, $(246^\circ\mbox{--}251^\circ,8.5^\circ\mbox{--}13.5^\circ)$, $(255^\circ\mbox{--}270^\circ,20^\circ\mbox{--}40^\circ)$,\\ $(268^\circ\mbox{--}271^\circ,46^\circ\mbox{--}49^\circ)$, and $(232^\circ\mbox{--}240^\circ,26^\circ\mbox{--}30^\circ)$.  Finally, we cut pixels that appeared to have severe photometric calibration errors ($\gtrsim5$ mag).  After these cuts, the survey region comprises 534,564 $N_{res}=9$ pixels covering $7010$ deg$^2$.

\subsubsection{SDSS LRGs}

We use the LRG catalog composed by \citet{2010ApJ...710.1444K}.  LRGs are the most luminous galaxies in the universe, making them important for probing large volumes.  They are also old stellar systems with uniform spectral energy distributions and a strong discontinuity at 4000 \AA, which enable precise photometric redshifts.  The LRG catalog consists of 105,623 spectroscopic LRGs from redshifts $0.16<z<0.47$.  We do not make any alterations to this catalog.  Our survey region comprises approximately 638,583 $N_{res}=9$ pixels covering $8374$ deg$^2$.

\subsection{WMAP-QSO Cross-Data Set}
We intersect the pixel sets from the temperature and quasar maps to produce two sets of maps that can be cross-correlated.  This operation leaves each map with 441,228 $N_{res}=9$ pixels covering 5786 deg$^2$ with $\sim$ 7 arcmin.~pixel resolution.  Since each WMAP band probes a separate redshift range for each of the CO emission lines, we also divide the quasar map into 8 maps, two for each of the WMAP bands.  Note that some of the bands for CO(1-0) will intersect in redshift with other bands for CO(2-1).  We list the properties of the 4 WMAP maps and 8 quasar maps in Table \ref{T:zbin} as well as the probed spatial scale determined by our pixel size.  Note that the Ka(2-1) redshift range exceeds the redshifts of the quasars; therefore, we will not determine any limits on the CO(2-1) line with the Ka band. 

\subsection{WMAP-LRG Cross-Data Set}
We also intersect the areas of the temperature and LRG maps for cross-correlation, with 619,708 $N_{res}=9$ pixels covering 8126 deg$^2$.  Note that the WMAP W band is the only band that overlaps with the LRG sample, and this is true only for the CO(1-0) line.  Thus, we will only get one constraint from this cross-correlation.  However, the number of LRGs in this redshift range is much more than the number of quasars, so we expect a more significant constraint than that from the quasars.

\begin{table*}[!t]
\begin{center}
\caption{\label{T:zbin} WMAP redshift bins for CO emission lines $J=1\to0$ and $J=2\to1$ with the quasar (QSO) counts from the QSO map (T x QSO intersected map). For reference, we write in parentheses the transverse scale corresponding to the pixel scale in Mpc/$h$ for each $z$ band probed. As stated above, the total number of QSOs in DR6 is 1,172,157.} 
\vspace{0.5cm}
\begin{tabular}{ccccc}
\hline
Band&$z(1\to0)$\quad ([Mpc/$h$])&$N_{\rm QSO}(1\to0)$&$z(2\to1)$\quad ([Mpc/$h$]) &$N_{\rm QSO}(2\to1)$ \\
\hline
Ka&2.151--2.898 (9) &90,780 (74,395)&5.302--6.796 (13) &80 (63)\\
Q&1.547--2.121 (7)  &146,111 (121,297)&4.094--5.242 (11) &573 (466)\\
V&0.691--1.130 (5)  &62,172 (51,818)&2.382--3.260 (9) &27,032 (22,180)\\
W&0.103--0.373 (1) &42,184 (34,400)&1.206--1.746 (6) &140,819 (116,791)\\
\hline
\end{tabular}\end{center}
\end{table*}

\section{Cross-Correlation Power Spectrum} \label{S:eqns}

We wish to correlate fluctuations in CO line emission with quasar and LRG maps.  Our CO temperature modeling will follow L11.  L11 derived the CO brightness temperature by calculating the specific intensity of CO emission as  a line-of-sight integral of the volume emissivity.  In L11, the star formation rate density (SFRD) used, given by their Eq.~6, is comparable with the value needed to reionize the universe at high $z$.  However, since we are interested in lower redshifts than in L11, the model must be modified, as we explain below.

\subsection{Model A: CO Luminosity -- Halo Mass} \label{S:modela}

The basic strategy is to construct a model that matches three
key observational inputs~\citep{2011ApJ...730L..30C}: the observed correlation between CO
luminosity and far-infrared luminosity, the far-infrared luminosity-star formation rate (SFR) correlation, and the observed SFRD of the Universe. In order to predict the spatial fluctuations in the
CO brightness temperature, we need to
further connect the star formation rate and host halo mass.

In comparison to the high redshift $z \sim 7$ case discussed in 
L11, this estimate should be on more solid
ground in several respects. The CO luminosity far-infrared
correlation and the correlation between the far-infrared luminosity and
star formation rate are measured at $z \sim 0-3$; their applicability
at higher redshift is questionable. In particular, the CO luminosity-SFR
correlation may drop at high redshift, since the low mass galaxies of
interest may have low metallicity, as well as an insufficient dust abundance to shield
CO from dissociating radiation. In addition, the increased CMB
temperature at high redshift may significantly reduce the contrast
between CO and the CMB.  Furthermore, the overall star formation
rate density is better determined at $z \sim 2$ than
at $z \sim 7$. On the other hand, the simplistic model adopted
in L11 to connect star formation rate to halo mass
is likely less applicable at low redshift, where various feedback effects
such as photoionization heating, supernovae, and AGN feedback have had more time
to operate. 

The starting point for the L11 model is the observed empirical
correlation between CO luminosity, $L_{\rm CO(1-0)}$, and far-infrared
luminosity, $L_{\rm FIR}$. Specifically, following L11 we use the
correlation reported in \citet{2010ApJ...714..699W}. Recent work, summarized in
\citet{Carilli:2013qm}, suggests however that rapidly star-forming
`starburst' galaxies may exhibit a somewhat different correlation
between CO and far-infrared luminosity than `main sequence' galaxies
with more gradual, yet longer lived star formation. In this work, for
simplicity, we ignore any trend in this correlation with galaxy
properties, and assume that the Wang et al. (2010) relation applies
globally.  Future measurements, further quantifying trends in this
relation with redshift, average galactic metallicity, and other galaxy
properties will help to refine our modeling. As in L11, we assume the
\citet{Kennicutt:1997ng} relation to connect far infrared luminosity and star
formation rate.

Lets turn to some quantitative estimates. The result of combining
the CO luminosity far-infrared correlation with the far-infrared SFR 
relation is, in L11,
\beqa
L_{\rm CO(1-0)} = 3.2 \times 10^4 L_\odot \left[\frac{\rm{SFR}}{M_\odot \rm{yr}^{-1}}\right]^{3/5}.
\label{eq:lco}
\eeqa
In L11 the authors further assumed a very simple model to convert SFR
to host halo mass:
\beqa 
\rm{SFR} &=& f_\star \frac{\Omega_b}{\Omega_m} \frac{M}{t_s} 
=  0.17 M_\odot \rm{yr}^{-1}\times \nonumber\\ && \left[\frac{f_\star}{0.1}\right] \left[\frac{\Omega_b/\Omega_m}{0.17}\right] \left[\frac{10^8 \rm{yrs}}{t_s}\right] 
\left[\frac{M}{10^9 M_\odot}\right].
\label{eq:sfr_mhost}
\eeqa
They took the fraction of baryons converted into stars to 
be $f_\star = 0.1$, and a (constant) star formation time scale of $t_s = 10^8$ yrs. 
Star formation was assumed to take place with equal efficiency 
in all halos above some some minimum host mass, $M_{\rm sfr, min}$.
It was found that the
implied star formation rate density is comparable to the critical
star formation rate density required to keep the Universe ionized, 
suggesting that this simple prescription is a reliable estimate. 
Furthermore, \citet{2007ApJ...668..627S} found that a similar model roughly
matches the luminosity function of Lyman Break Galaxies (LGBs) at $z=6$.
Unfortunately, this simple prescription will not provide a good
estimate of the SFR at lower redshifts, as we discuss.

We now explain why the estimate for $\VEV{T_{\rm CO}}$ in L11 cannot be extrapolated to low redshifts.
Taken together, Equations \ref{eq:lco} and \ref{eq:sfr_mhost} imply
that $L_{\rm CO(1-0)} \propto M^{3/5}$. For simplicity, we assumed that
$L_{\rm CO(1-0)}$ instead {\em scaled linearly with halo mass} after matching the implied
CO luminosity at $M=10^8 M_\odot$. Since massive halos are rare at $z \sim 7$,
the CO emissivity in this model is dominated by galaxies in low mass halos close
to the minimum CO luminous halo mass, $M_{\rm co, min}$,
and so adopting a linear scaling rather than a $M^{3/5}$
power law, has relatively little impact on the mean CO brightness temperature.
At $z \sim 7$ the linear scaling leads to only a slight overestimate compared
to the $M^{3/5}$ power law.
Together, these assumptions implied $L_{\rm CO(1-0)}(M) = 2.8 \times 10^3 L_\odot
M/(10^8 M_\odot)$.  Note that \citet{2011ApJ...741...70L} focused
on CO(2-1) assuming that the $J=2-1$ and $J=1-0$ lines have the 
same luminosity, which
is very conservative. In the optically thick, high temperature limit, 
the brightness temperature
would be a factor of $8$ larger. At $z \sim 7$ this more than makes up
for the possible overestimate from the linear scaling.

Now it is easy to see that applying this blindly at $z \sim 2$ may
lead to problems because significantly more massive halos are no longer
rare.
As a result, if we assume a linear scaling when
the true scaling is sub-linear, we will significantly overestimate the
CO emissivity and brightness temperature.
For example, the above $L_{\rm CO(1-0)} - M$ relation gives
$L_{\rm CO}(M = 10^{11} M_\odot) = 2.8 \times 10^6 L_\odot$. However if
we had used the sublinear scaling, we would have obtained
$L_{\rm CO}(M = 10^{11} M_\odot) = 1.8 \times 10^5 L_\odot$, which is lower by a factor of $\sim 15$.
Although the CO luminosity halo mass relation used in L11
seems like a plausible estimate at high redshifts, where low mass galaxies
dominate the SFRD and CO emissivity, blindly extrapolating it to 
higher halo masses
is problematic.

We derive a more accurate expression by first adjusting the SFR-M prescription to be
more suitable at low redshifts, where feedback processes will further
suppress star formation in low mass halos. 
We follow \citet{2003ApJ...586..693W} in
adopting a halo-mass dependent star formation efficiency, as suggested
by the $z \sim 0$ observations of \citet{2003MNRAS.341...54K}. In particular,
we assume that the fraction of gas turned into stars above $M_{\rm sfr, min}$ 
scales
as $M^{2/3}$ below some mass $M_0$, and that the star formation efficiency
is independent of mass above the characteristic mass $M_0$. More specifically we assume 
\beqa
SFR = && f_\star \frac{\Omega_b}{\Omega_m} \frac{M_o}{t_s}
\left[\frac{M}{M_0}\right]^{5/3}; M \leq M_0 \nonumber \\
SFR = && f_\star \frac{\Omega_b}{\Omega_m} \frac{M}{t_s}; M \geq M_0.
\label{eq:sfrnew}
\eeqa
In reality $f_\star$, $M_o$ and $t_s$ likely have some redshift dependence.
Here we fix the characteristic mass scale $M_0$ at $M_0 = 5 \times 10^{11} M_\odot$ (close to the mass scale at $z \sim 2$ in the \citet{2003ApJ...586..693W} model),
and normalize the proportionality constant $f_\star \Omega_b/\Omega_m$
to match the observed SFRD at $z \sim 2$, $\dot{\rho_\star} = 0.1 M_\odot$ yr$^{-1}$ Mpc$^{-3}$ \citep{2006ApJ...651..142H}. The mass dependent efficiency reduces the efficiency of star
formation in low mass halos, reflecting the impact of feedback processes.
It is convenient that the SFR scales as $M^{5/3}$ in this model, because
this yields a linear scaling of $L_{\rm CO} \propto M$, {\em although with
significantly lower normalization than the previous relation}. (This is strictly
true only below $M_0$ but these halos dominate the CO emissivity even
at low $z$ and so it is safe to adopt this scaling for all halo
masses here.) This means that the only
thing that changes in this model is the brightness temperature normalization $\VEV{T_{\rm CO}}$,
and not the bias and Poisson terms.

Combining the revised SFR model with Equation \ref{eq:lco}
we find:
\beqa 
L_{\rm CO(1-0)} = 1 \times 10^6 L_\odot \left[\frac{M}{5 \times 10^{11} M_\odot}\right].
\label{eq:lco_fix}
\eeqa
This normalization is a factor of $\sim 14$ lower than in L11
but is likely a better match at $z \sim 2$ where star formation occurs
mostly in substantially more massive halos than at $z \sim 7$.

\begin{figure}[!t]
\begin{center}
\includegraphics[width=0.5\textwidth]{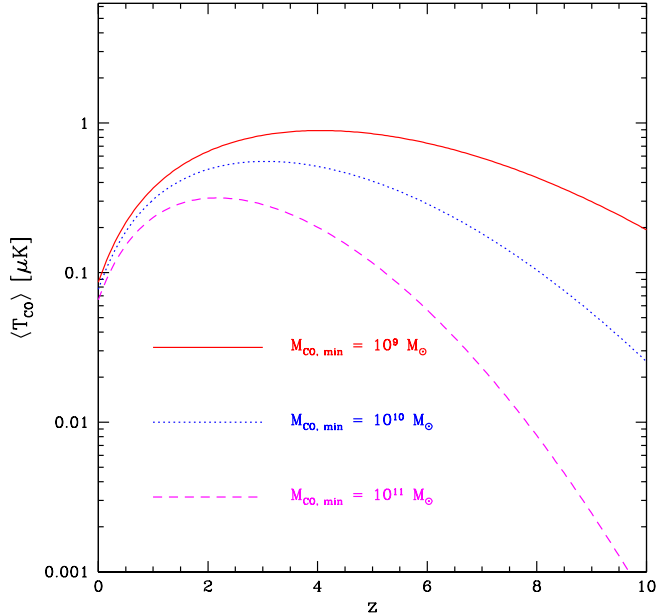}
\caption{\label{F:tco} The mean CO brightness temperature in Model A as a function of redshift.  The solid, dotted, and dashed lines denote $\VEV{T_{\rm CO}}$ for $M_{\rm co,min}=$10$^9$, 10$^{10}$, and 10$^{11}M_\odot$, respectively. Note the same mean $\VEV{T_{\rm CO}}$ is assumed for both the CO(1-0) and CO(2-1) lines.}
\end{center}
\end{figure}

In order to turn this into an expression for the brightness temperature
we can follow the formulas from L11, but with
small modifications. First, we will assume the optically thick, high-temperature limit. This should be a good approximation for the
$J=1-0$ and $2-1$ lines at the redshifts of interest. We assume that
the duty cycle of CO luminous activity matches the star formation duty
cycle, i.e., we assume $f_{\rm duty} = t_s/t_{\rm age}(z)$, where we
fix $t_s \sim 10^8$ yrs, and $t_{\rm age}(z)$ is the age of the Universe.
The result of this calculation with the normalization of
Eq.~\ref{eq:lco_fix} is
\beqa
\VEV{T_{\rm CO}}(z_J) =  0.65 \mu K \left[\frac{f_{\rm coll}(M_{\rm co, min}; z_J)}{0.3}\right] \left[\frac{3.4 \times 10^9 \rm{yr}}{t_{\rm age}(z)}\right] \times&&\nonumber\\ 
\left[\frac{H(z_J=2)}{H(z_J)}\right] 
\left[\frac{1+z_J}{3}\right]^2,\,\,\,\,\,\,\,&&
\label{eq:tco_lowz}
\eeqa
where we have scaled to characteristic values at $z \sim 2$.
Note that the redshift scaling is $\propto (1+z_J)^{1/2}$ only
in the high redshift limit where $H(z_J) \propto (1+z_J)^{3/2}$. 
The results are plotted in Figure \ref{F:tco}. In this model
the star formation efficiency declines below the characteristic mass scale
$M_0 \sim 5 \times 10^{11} M_\odot$,
which may underestimate the higher redshift signal if low mass galaxies
form stars efficiently at high redshift and are CO luminous. 
However, our present
data sets provide little constraint at very high redshift, and so we
are less concerned with the predictions there.  Note that $\VEV{T_{\rm CO}}$ in this model is $0.1-1\mu$K for all $z$ (for $M_{\rm co, min}=10^9M_\odot$) and thus will be difficult to see in a CMB experiment like WMAP.

\subsection{Model B: CO Luminosity -- Star Formation Rate}
We recognize that a weak part of the above estimate is the simplistic model
connecting SFR and host halo mass. In this section we take
a more empirical approach to estimating the spatially-averaged
CO brightness temperature and thereby circumvent the need to connect
SFR and host halo mass. In particular, here we use
recent determinations of the star formation rate (SFR) function from \citet{2012ApJ...756...14S}.
These authors use measurements of the UV luminosity function along with
the Kennicutt \citep{1998ARA&A..36..189K} relation, connecting SFR and UV luminosity to determine
the SFR function, i.e., the number density of galaxies with a star formation rate between SFR and SFR + dSFR.
The observed UV luminosity is corrected for dust attenuation based on the slope of the UV continuum
spectra, which gives a luminosity and redshift-dependent extinction correction.
They fit Schechter functions to the resulting SFR functions using new UV luminosity function and extinction 
measurements at $z \sim 4-7$, and tabulate results from the literature at lower redshift (their Table 3).

The Schechter function \citep{1976ApJ...203..297S} is:
\beqa
\Phi(SFR)dSFR = \phi_\star \left(\frac{SFR}{SFR_\star}\right)^\alpha \rm{exp}\left[-\frac{SFR}{SFR_\star}\right] \frac{dSFR}{SFR_\star},
\label{eq:phi_sfr}
\eeqa
and is parameterized by a characteristic $SFR$, $SFR_\star$, a characteristic number density, $\phi_\star$, and
a faint-end slope, $\alpha$. The star formation rate density, $\dot{\rho_\star}$, can be derived by integrating
the SFR over the Schechter function. Assuming that the Schechter function form is maintained to arbitrarily low
SFRs then gives:
\beqa
\dot{\rho_\star} = \phi_\star SFR_\star \Gamma[2 + \alpha].
\label{eq:sfrd}
\eeqa
Here $\Gamma[2+\alpha]$ is a Gamma function.

We can further combine the SFR Schechter function with the observed $L_{\rm CO (1-0)} - SFR$ correlation (Equation \ref{eq:lco}), to
estimate the co-moving (frequency-integrated) CO emissivity. As discussed below, we will assume this prescription only for
sources above some minimum SFR, $SFR_{\rm min}$, and that sources below this critical SFR do not contribute
appreciably to the CO emissivity.  Combining Equation \ref{eq:lco} and Equation \ref{eq:phi_sfr}, the resulting
co-moving emissivity is:
\beqa
\epsilon_{\rm CO (1-0)} = \phi_\star L_0 \left(\frac{SFR_\star}{1 M_\odot/yr}\right)^{3/5} \Gamma\left[\alpha + 1.6, \frac{SFR_{\rm min}}{SFR_\star}\right].
\label{eq:eps_co}
\eeqa
In this equation, $L_0$ denotes the luminosity in the CO(1-0) line for a SFR of $SFR= 1 M_\odot/$yr. We fix this
at $L_0 = 3.2 \times 10^4 L_\odot$ as in Equation \ref{eq:lco}, and as suggested by $z \sim 0-3$ CO observations. The factor
$\Gamma[\alpha+1.6, SFR/SFR_\star]$ is an Incomplete Gamma Function. This reveals the importance
of the sub-linear scaling of Equation \ref{eq:lco} for the CO emissivity: combining the sub-linear scaling
and the Schechter form for the SFR function yields a formally divergent CO emissivity as $SFR \rightarrow 0$ for
$\alpha \leq -1.6$. This may be an artifact of extrapolating the Schechter form and/or the sub-linear
$L_{\rm CO (1-0)}$ scaling with SFR to arbitrarily low SFR; presumably one or both of these relations drop-off at low SFR. For instance, low luminosity
dwarf galaxies may have small metallicites and fall below the extrapolation
of the $L_{\rm CO}-SFR$ relation.  These relations generalize the approach
of \citet{2011ApJ...730L..30C}, who estimated the CO emissivity from the SFRD assuming the two scale linearly with SFR.
Our more accurate relation requires, however, specifying $SFR_{\rm min}$ to relate the co-moving CO emissivity and the SFRD.

Using the Equations in L11 we can relate the co-moving CO emissivity as calculated above to the spatially-averaged
CO brightness temperature. It is useful to note that this gives:
\beqa
\VEV{T_{\rm CO}}(z_J) = 0.65 \mu K \left[\frac{\epsilon_{\rm CO (1-0)}}{6.3 \times 10^2 L_\odot \rm{Mpc}^{-3}}\right]\times&&\nonumber\\
\left[\frac{H(z_J=2)}{H(z_J)}\right] 
\left[\frac{1+z_J}{3}\right]^2.\,\,\,\,\,\,\,&&
\label{eq:tco_eps}
\eeqa

\begin{figure}
\begin{center}
\includegraphics[width=0.5\textwidth]{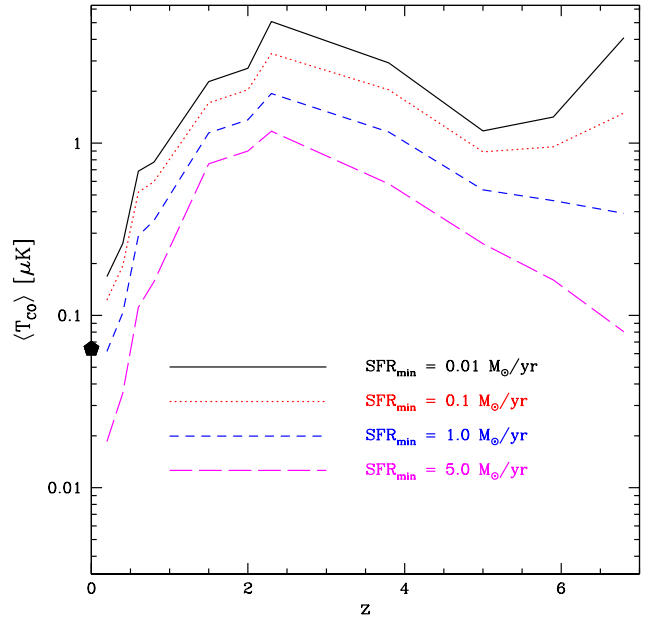}
\caption{The average CO brightness temperature in Model B. These results are estimated from
Equations \ref{eq:eps_co} and \ref{eq:tco_eps} using the SFR function fits from \citet{2012ApJ...756...14S}. The
solid pentagon at $z=0$ is the average CO brightness temperature inferred from the CO luminosity function
measured at $z=0$ by \citet{2003ApJ...582..659K}.}
\label{fig:tco_emp}
\end{center}
\end{figure}

We calculate the results of Equations \ref{eq:eps_co} and \ref{eq:tco_eps} from $z=0.2-6.8$ using the best-fit results
in Table 3 of \citet{2012ApJ...756...14S}. We focus on the UV- and MIR-derived SFR functions from that paper,  and do not
include the H$\alpha$-based estimates, although these appear to be comparable. We ignore the point at $z=0$
in their Table 3 from \citet{2011MNRAS.415.1815B}, since there appears to be a typo in the fit parameters from this paper, 
as remarked in the footnote to Table 3.
The results of these calculations are shown in Fig.~\ref{fig:tco_emp}. The overall redshift evolution shown here is
noisy because $\VEV{T_{\rm CO}}$ is estimated from several discrete observational measurements, each with significant
observational error bars.
As anticipated above,
the brightness temperature estimates are sensitive to $SFR_{\rm min}$, which is somewhat uncertain. The observed
$L_{\rm CO (1-0)}-SFR$ correlation includes galaxies with SFRs down to $\sim 0.5 M_\odot/$yr~\citep{2010ApJ...714..699W}, but
the data are a bit sparse at low SFR. The Schechter function fits probe galaxies down to typical SFRs of $\sim0.1-1 M_\odot/$yr, with some redshift dependence at the faint-end limit (\textit{e.g.}~Fig.~2 of \citet{2012ApJ...756...14S}).
The results are
particularly sensitive to $SFR_{\rm min}$ at $z = 6.8$, where the slope of the faint-end luminosity 
function is especially steep ($\alpha = -1.96$).
The more empirical model shown here agrees with the models of the previous section of Figure \ref{F:tco} at the order-of-magnitude level,
although the results in the $SFR_{\rm min} = 0.01 M_\odot/$yr case are larger by a factor of $\sim 4$ at $z \sim 2$
than the earlier $M_{\rm CO min} = 10^9 M_\odot/$yr model. The high-redshift results at $z \gtrsim 5$ are also broadly 
consistent with the estimates in L11.
There are still significant uncertainties in the
Schechter function fits to the SFRs, and in the extrapolations to the faint end, but this latter estimate
is probably more secure than the estimates of the previous section which rely on an oversimplified model to connect SFR and halo mass.  Note that we will estimate power spectra for Model B by multiplying the Model A spectra by $\VEV{T_{\rm CO}}^2({\rm Model\,B})/\VEV{T_{\rm CO}}^2({\rm Model\,A})$, although strictly speaking this is not accurate since the clustering bias depends on the SFR-M relation, which is not determined for Model B.

As one final sanity check, we can use the measured CO(1-0) luminosity function from \citet{2003ApJ...582..659K} at $z=0$ to estimate the CO emissivity
and brightness temperature at $z=0$. They provide a Schechter function fit with $\rho(L) = d \Phi/d\log_{\rm 10} L
= \rho_\star \ln(10) (L/L_\star)^{\alpha+1} \exp(-L/L_\star)$, and $\rho_\star = 7.2 \times 10^{-4}$ Mpc$^{-3}$ mag$^{-1}$,
$\alpha = -1.30$. \citet{2003ApJ...582..659K} quotes results in terms of velocity-integrated CO luminosities, with a characteristic $L_\star$
value of
$L_{\rm CO, V; \star} = 1.0 \times 10^7$ Jy km s$^{-1}$ Mpc$^2$. We convert this into solar units using
the relations in the Appendix of \citet{2009ApJ...702.1321O}, finding $L_\star = 9.6 \times 10^4 L_\odot$. Using the best-fit Schechter function parameters from \citet{2003ApJ...582..659K}, and integrating to $L_{\rm CO(1-0)} = 0$ gives $\VEV{T_{\rm CO}} = 0.064 \mu K$, broadly consistent with our estimates, extrapolated to $z=0$ (see the solid pentagon in Fig. \ref{fig:tco_emp}). 

\subsection{CO Clustering}
\begin{figure}
\begin{center}
\includegraphics[width=0.45\textwidth]{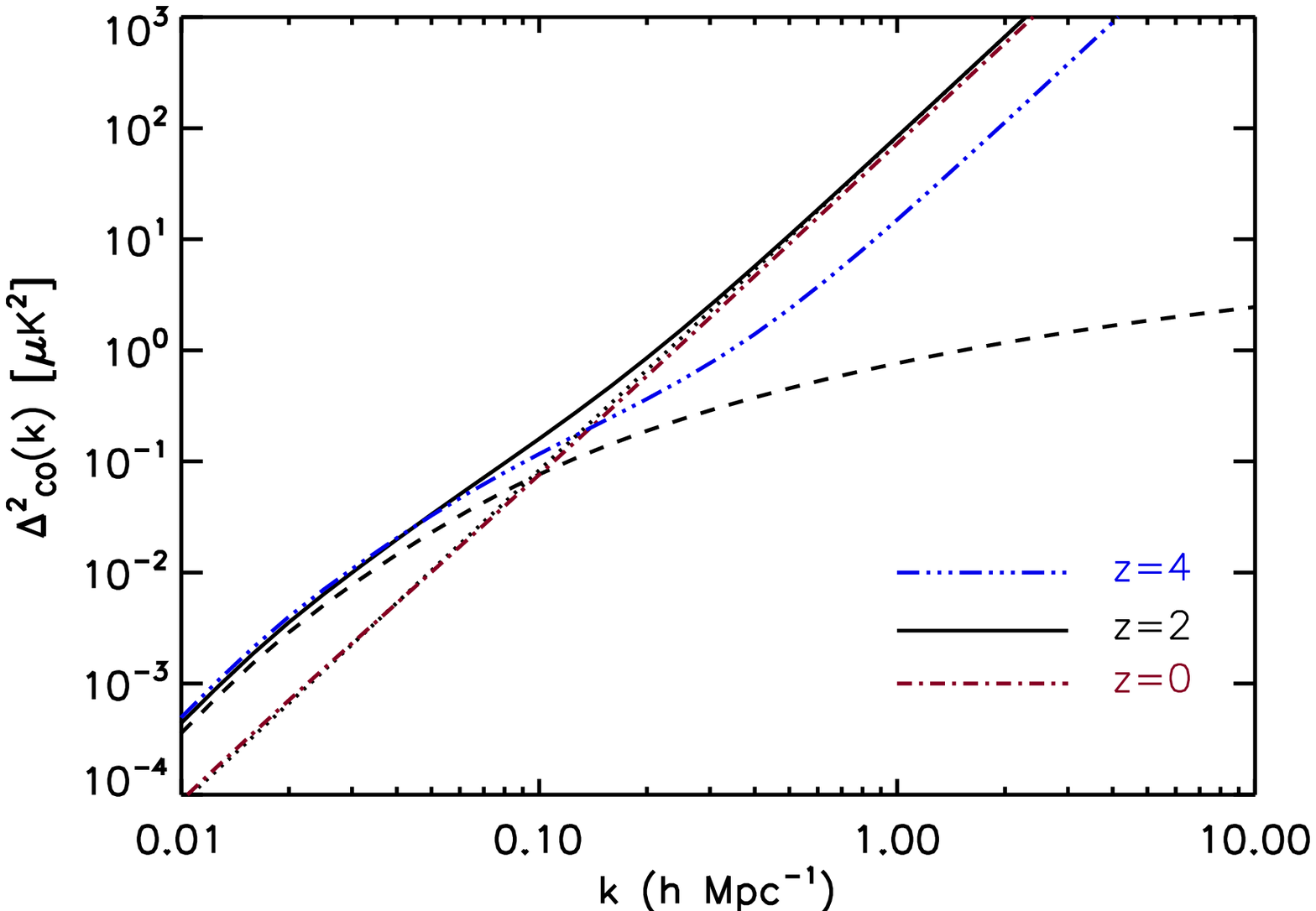}
\includegraphics[width=0.45\textwidth]{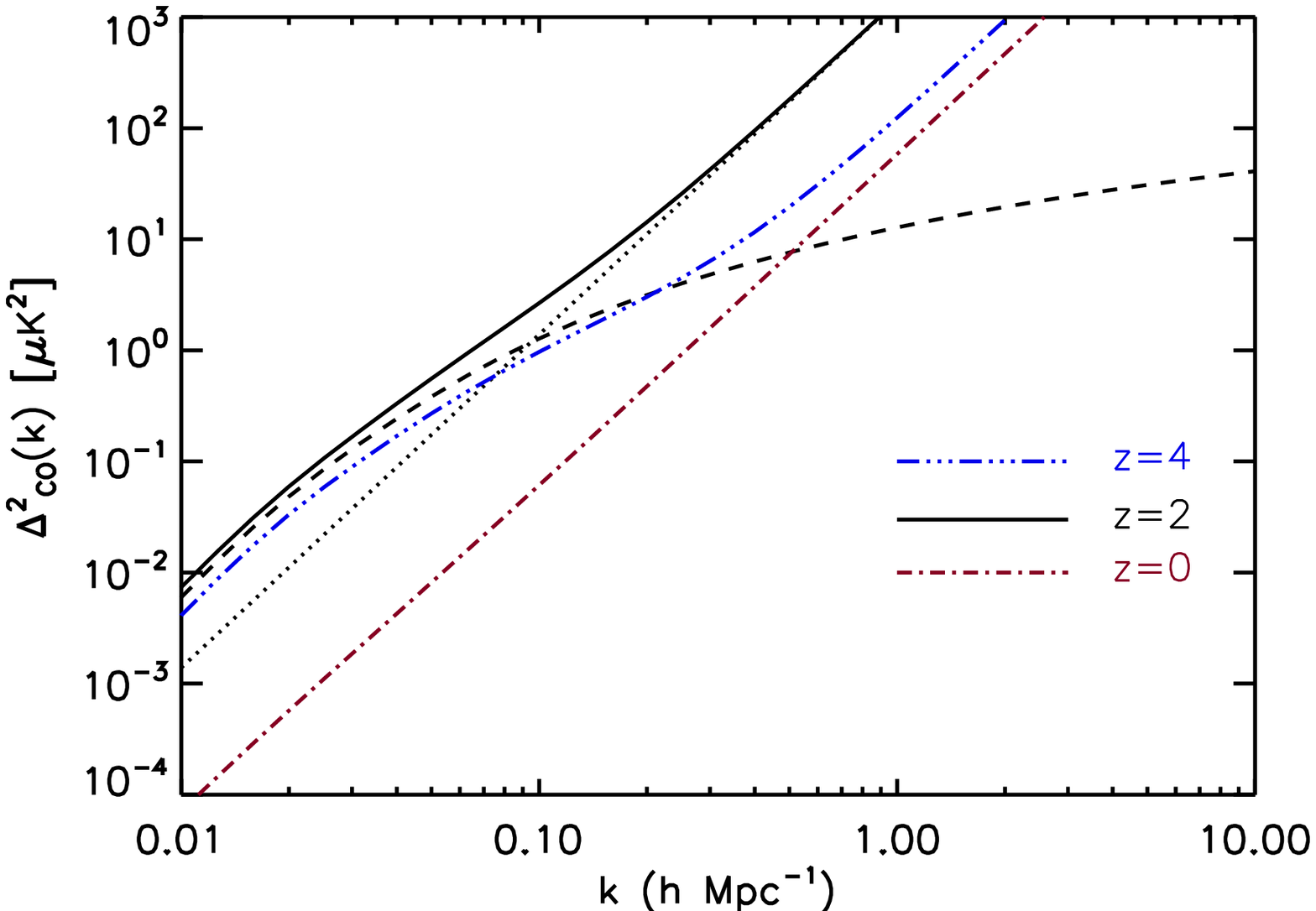}
\caption{The 3D power spectrum for CO fluctuations in the redshift range $0<z<4$.  The top panel is for Model A with $M_{\rm co,min}=10^9M_\odot$, while the bottom panel is for Model B with $SFR_{\rm min}=0.01M_\odot/$yr.  The black dashed (dotted) line shows the clustering (shot noise) term for $z=2$.}
\label{F:pk}
\end{center}
\end{figure}

In order to model the clustering of CO emitters, following L11, we will assume the standard relation between CO luminosity and host halo mass, linear bias, that the scales of interest are much larger than the virial radius of the relevant halos so that we can neglect the non-linear term, and that the halo shot noise obeys Poisson statistics. With this assumption, the 3D power spectra for CO temperature fluctuations goes as 
\begin{eqnarray}
\label{Eq:Pco}
P_{\rm CO}(k,z) &=& \VEV{T_{\rm CO}}^2(z)\left[b^2(z)P_{\rm lin}(k,z)+\frac{1}{f_{\rm duty}(z)}\frac{\VEV{M^2}}{\VEV{M}^2}\right]\nonumber\\
\end{eqnarray}
where $b(z)$ is the effective $z$-dependent halo bias given in Eq.~15 of L11
\begin{eqnarray}
b(z)=\frac{\int_{M_{\rm co,min}}^\infty dM\,M\frac{dn}{dM}b(M,z)}{\int_{M_{\rm co,min}}^\infty dM\,M\frac{dn}{dM}}\, ,
\end{eqnarray}
where $dn/dM$ is the halo mass function from \citet{2008ApJ...688..709T} and $M$ is the associated halo mass, and $P_{\rm lin}(k,z)$ is the linear matter power spectrum. The second term in brackets for $P_{\rm CO}$ is the shot noise, with
\begin{eqnarray}
\VEV{M^2}&=&\int_{M_{\rm co,min}}^\infty dM\,M^2\frac{dn}{dM}\, ,\nonumber\\
\VEV{M}&=&\int_{M_{\rm co,min}}^\infty dM\,M\frac{dn}{dM}\, .
\end{eqnarray}
We implicitly assume that every dark matter halo will host at least one CO emitter with a duty cycle $f_{\rm duty}(z)$.  We plot the CO three-dimensional power spectrum in Fig.~\ref{F:pk}.

We also model the clustering of quasars and LRGs.  We assume the quasars, LRGs, and CO emitters trace the same LSS such that cross-correlating the temperature maps with either the quasars or LRGs will uncover clustering of CO emitters.  Note, however, that we do not require nor assume the quasars or LRGs produce CO emission.  The objective of this paper is not to find all CO gas; it is in fact to use halos with CO gas to trace LSS.  Since the amount of CO gas in halos should not statistically vary across the sky, CO emission should be a good tracer of LSS, or the density distribution of massive halos, across the sky. Thus, we do not need to claim that quasars trace the CO gas. We only need the quasars to trace LSS and the CO emitters to trace LSS.

We also assume linear biasing so that the 3D quasar and LRG spectra are
\begin{eqnarray}
 P_{Q}(k,z) &=&b_Q^2(z)P_{\rm lin}(k,z)\nonumber\\
 P_{LRG}(k,z) &=&b_{LRG}^2P_{\rm lin}(k,z)\, ,
\end{eqnarray}
where $b_Q(z)$ is the quasar clustering bias that we compute using the estimate of Eq.~15 in \citet{2005MNRAS.356..415C} and $b_{LRG}$ is the LRG clustering bias.  In general, the LRG bias would be redshift-dependent, but we will assume a constant bias for the LRGs.  Similarly, the 3D cross power-spectra can be written as 
\begin{eqnarray}
 P_{CO-Q}(k,z) &=&r_Q\VEV{T_{\rm CO}}(z)b(z)b_Q(z)P_{\rm lin}(k,z)\nonumber\\
 P_{CO-LRG}(k,z) &=&r_{LRG}\VEV{T_{\rm CO}}(z)b(z)b_{LRG}P_{\rm lin}(k,z)\, ,
\end{eqnarray}
where we include cross-correlation coefficients $r_Q$ and $r_{LRG}$ between CO emitters and quasars and LRGs, respectively, which we assume to be scale- and redshift-independent throughout this analysis.  Note it is possible that quasars and LRGs may live in the same halos as CO emitters, contributing a shot noise term to their cross-correlation.  We will neglect this shot noise in the analysis.

Assuming the Limber approximation in the small scale limit (typically $\ell>10$), we can then derive the angular auto/cross power spectra. The cross-power spectra have the form
\begin{eqnarray}
\label{Eq:ClCOQ}
C_\ell^{CO-Q}&=&\int dz \frac{H(z)}{c}\frac{f_{\rm CO}(z)f_Q(z)}{\chi^2(z)}\times\nonumber\\&&P_{CO-Q}[k=\ell/\chi(z),z]\nonumber\\
C_\ell^{CO-LRG}&=&\int dz \frac{H(z)}{c}\frac{f_{\rm CO}(z)f_{LRG}(z)}{\chi^2(z)}\times\nonumber\\
&&P_{CO-LRG}[k=\ell/\chi(z),z]\, ,
\end{eqnarray}
where $H(z)$ is the Hubble parameter and $f_{\rm CO}(z)$, $f_Q(z)$, and $f_{LRG}(z)$ are selection functions of the CO temperature fluctuations, the quasar distribution, and the LRG distribution, respectively. In our analysis, we assume flat, normalized selection functions for CO, quasars, and LRGs as a rough estimate. The three relevant angular auto-power spectra are given by
\begin{eqnarray}\label{E:clauto}
C_\ell^{\rm CO}&=&\int dz \frac{H(z)}{c}\frac{f_{\rm CO}^2(z)}{\chi^2(z)}P_{\rm CO}[k=\ell/\chi(z),z]\nonumber\\
C_\ell^{Q}&=&\int dz \frac{H(z)}{c}\frac{f_{Q}^2(z)}{\chi^2(z)}P_{Q}[k=\ell/\chi(z),z]\nonumber\\
C_\ell^{LRG}&=&\int dz \frac{H(z)}{c}\frac{f_{LRG}^2(z)}{\chi^2(z)}P_{LRG}[k=\ell/\chi(z),z]\, .
\end{eqnarray}

\begin{figure}
\begin{center}
\includegraphics[width=0.45\textwidth]{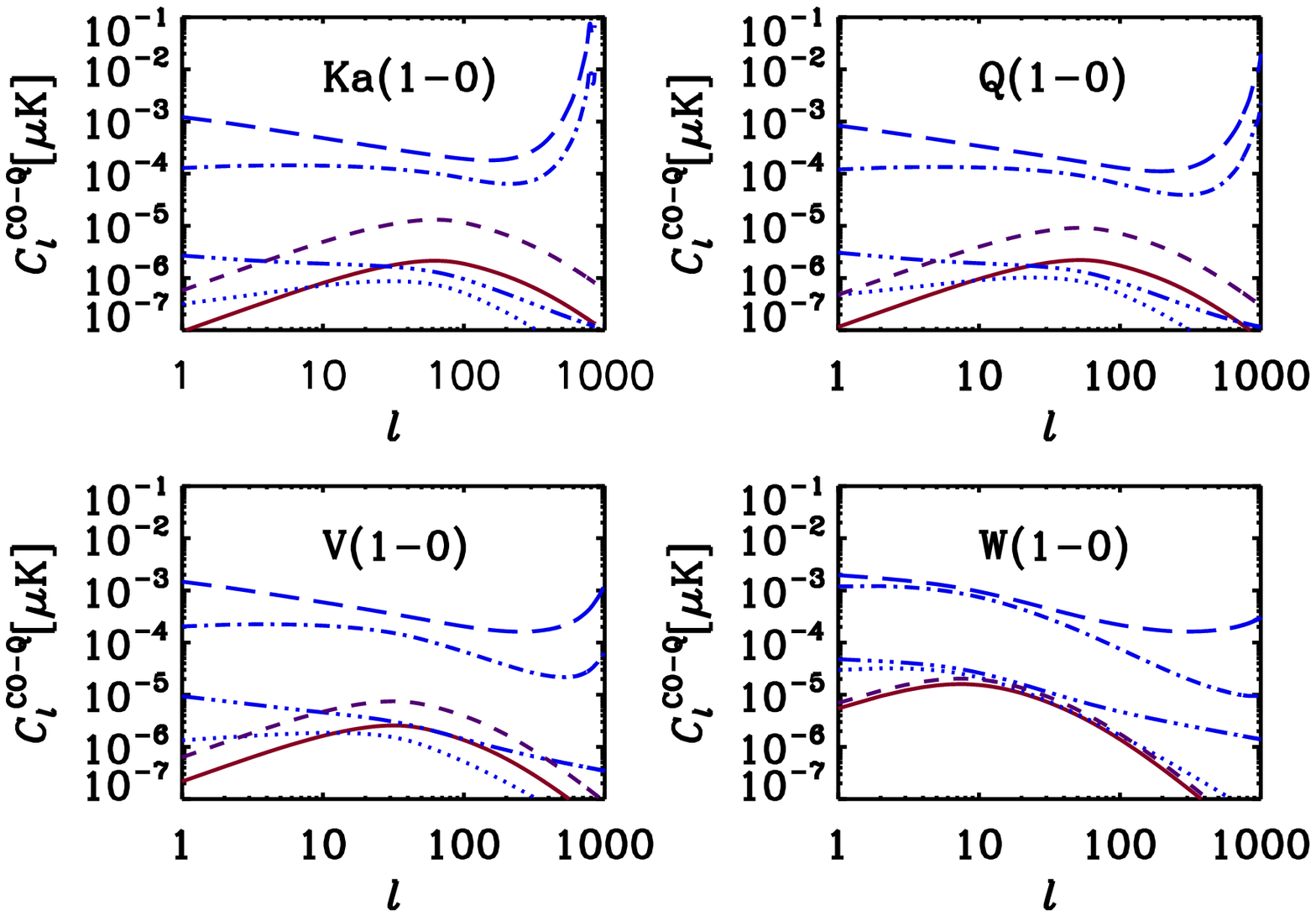}
\includegraphics[width=0.45\textwidth]{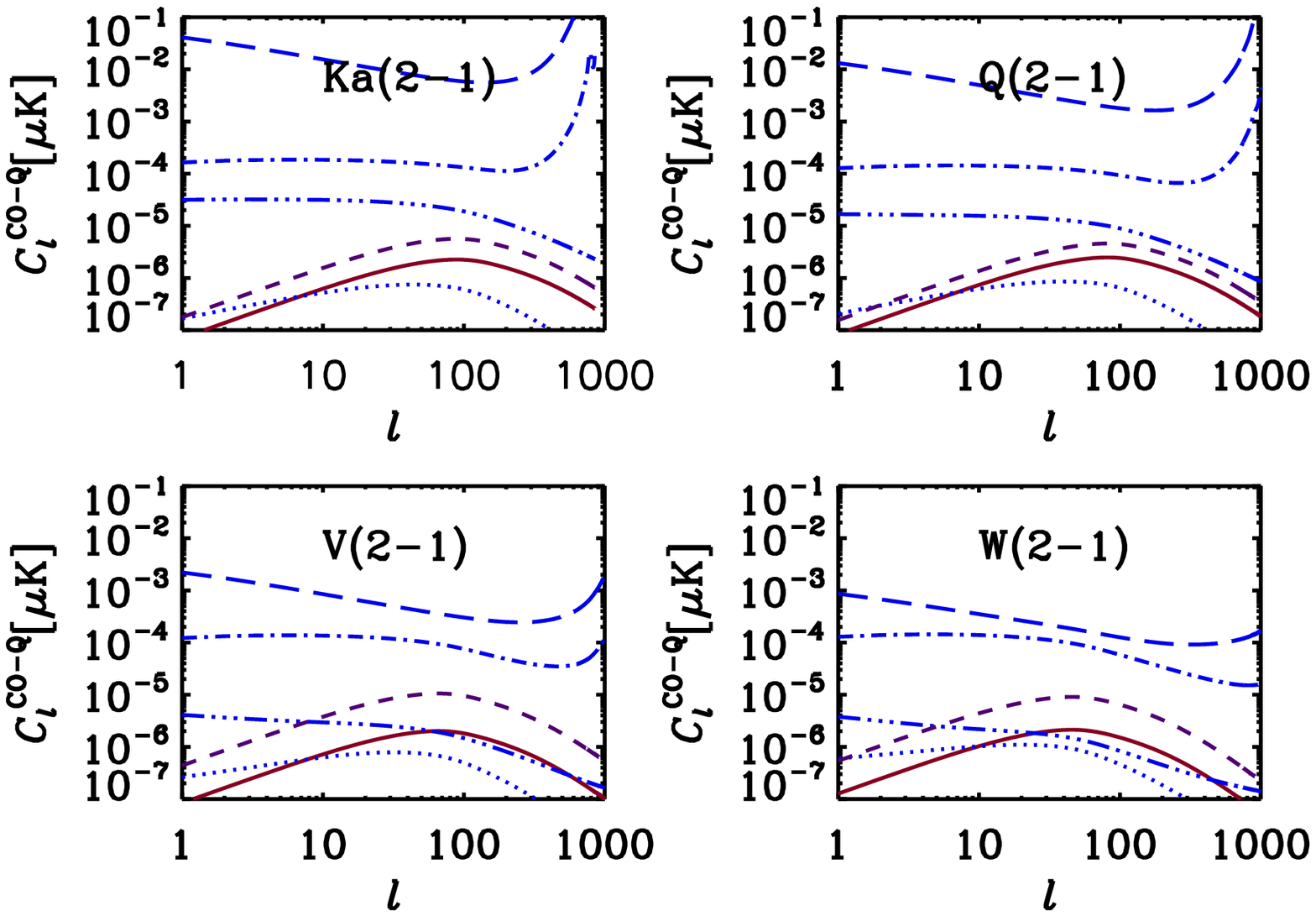}
\caption{\label{F:cln} The predicted angular cross-power spectra and noise curves with each WMAP band cross-correlated with its corresponding quasar map.  The solid, red line is the cross-correlation signal for CO line emission for Model A with $M_{\rm co,min}=10^9M_\odot$, and the purple, short-dashed line is for Model B with $SFR_{\rm min}=0.01M_\odot/$yr. The blue lines are the noise curves assuming Model A for the following cases: cosmic variance and CO shot noise limited (dotted), including WMAP instrumental noise (dot-dashed) only, including quasar shot noise only (dot-dot-dot-dashed), and including both noise sources (dashed).  Note that the noise curves are slightly modified for Model B.}
\end{center}
\end{figure}

\begin{figure}
\begin{center}
\includegraphics[width=0.45\textwidth]{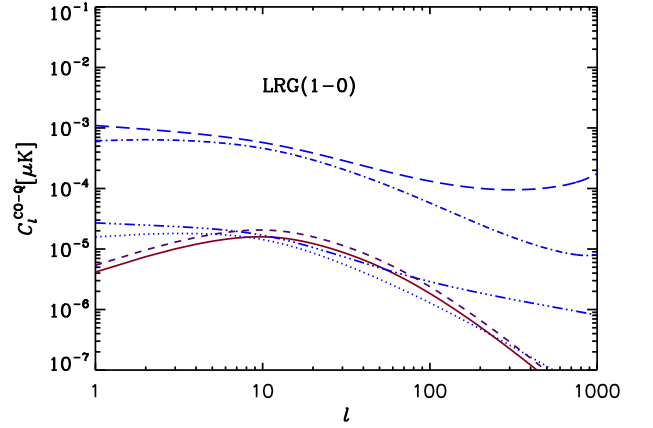}
\caption{\label{F:clnlrg} The predicted angular cross-power spectrum and noise curves for Model A with the WMAP W band cross-correlated with the LRG map for $M_{\rm co,min}=10^9M_\odot$.  The solid, red line is the cross-correlation signal for CO line emission, and the purple, short-dashed line is for Model B with $SFR_{\rm min}=0.01M_\odot/$yr.  The blue lines are the noise curves for the following cases: cosmic variance and CO shot noise limited (dotted), including WMAP instrumental noise (dot-dashed) only, including quasar shot noise only (dot-dot-dot-dashed), and including all noise (dashed).  Note that the noise curves would be slightly modified for Model B.}
\end{center}
\end{figure}

We plot the predicted angular power cross-spectra for Model A with quasars in Fig.~\ref{F:cln} and the cross-spectra with LRGs for $M_{\rm co,min}=10^9M_\odot$ in Fig.~\ref{F:clnlrg}.  On the plot we superimposed an estimate for Model B by just rescaling $\VEV{T_{\rm CO}}$.  For the LRGs we assume $b_{LRG}=2.48$, which we measured from the data (see Sec.~\ref{S:lrgresult}). Given the finite redshift widths defined by the WMAP bands, the evolution of the peaks of the angular spectra with the band frequency reflects the evolution in angular diameter distance with redshift. 
A higher band frequency means a lower redshift, which means a larger angular scale (lower $\ell$) for the peak.  For each band, the CO(1-0) spectrum can be higher or lower than the CO(2-1) spectrum depending on the redshift ranges probed, which determine the halo, quasar, and LRG biases and the CO brightness temperature.

Given Model A, we also find that $C_{\ell=200}^{CO(1-0)}$ ($C_{\ell=200}^{CO(2-1)}$)= 3.90$\times10^{-7}$ (2.64$\times10^{-7}$), 3.48$\times10^{-7}$ (3.88$\times10^{-7}$), 1.93$\times10^{-7}$ (3.73$\times10^{-7}$), and 7.80$\times10^{-8}$ (2.83$\times10^{-7}$) $\mu$K$^2$ respectively in the Ka, Q, V and W bands if we neglect the shot noise term, which becomes dominant only at $\ell$ = 579 (N/A), 292 (N/A), all $\ell$ (744), and all $\ell$ (170) for the same bands. Note that these numbers change if $M_{\rm co,min}\neq10^9M_\odot$.  For reference, the CMB angular power spectra is $C_\ell^{CMB}\simeq$ 1 $\mu$K$^2$ at $\ell$=200 so it will act as an important source of noise, if not subtracted.  
High quasar shot noise causes CO and quasars to not be perfectly correlated even on the largest scales. We note that the shot noise is more important at lower $z$ (lower $\nu$) as the number of massive halos increase more rapidly than the number of small mass halos. As will be discussed below, the instrumental noise at $\ell=200$ is 0.0253, 0.0230, 0.0347, and 0.0501 $\mu$K$^2$.  In Figs.~\ref{F:cln} and \ref{F:clnlrg}, we also present the various sources of noise in comparison to the cross-correlation signal.  Note that we do not include foreground noise, although it is included in the final results.  Other than foreground noise, we see that the spectra are dominated by the WMAP instrumental noise.

Given these numbers, the prospect of measuring a cross-correlation signal are not very high given the current state of our model. However, given the theoretical uncertainties, we will still attempt to measure directly this correlation with currently available data.

\section{Measurement Methodology}\label{S:method}

We estimate the CO--LSS tracer angular power spectrum, where the LSS tracer can be either quasars or LRGs, using a minimum-variance estimator of the form
\begin{eqnarray}\label{E:clest}
\hat{C}_\ell^{CO-Tr}=\sum_{m=-\ell}^\ell \frac{\bar{a}_{\ell m}^{\rm CO}\bar{a}_{\ell m}^{Tr*}}{(2\ell+1)f_{\rm sky,CO-Tr}}\, ,
\end{eqnarray}
where $f_{\rm sky,CO-Q}=0.140$ and $f_{\rm sky,CO-LRG}=0.197$ are the sky coverage fraction of the WMAPxSDSS map intersections for quasars and LRGs, respectively, and $\bar{a}_{\ell m}^X$ are the inverse-noise-filtered spherical harmonic coefficients for observable $X$, given by
\begin{eqnarray}\label{E:invfilt}
\bar{a}_{\ell m}^X=(C_\ell^X+C_\ell^{n,X})[(\mathbf{S}_X+\mathbf{N}_X)^{-1}\mathbf{d}_X]_{\ell m}\, ,
\end{eqnarray}
where $C_\ell^X$ and $C_\ell^{n,X}$ are the fiducial signal and noise angular auto-power spectra, respectively.  We will assume Model A in Sec.~\ref{S:modela} with $M_{\rm co,min}=10^9M_\odot$ for the fiducial signal auto-power spectrum, noting that the estimator should not depend significantly on the fiducial spectrum.  The CO thermal noise and quasar shot noise angular power spectra are 
\begin{equation}\label{E:clnco}
C_\ell^{n,CO_i}=\Delta\Omega\sigma_{T_{ij}}^2/W_\ell^{T_i}\, ,
\end{equation}
and 
\begin{equation}\label{E:shot}
C_\ell^{n,Tr_i}=(\bar{n}_{i_{tr}})^{-1}\, ,
\end{equation}
for the $i$ band, where $j$ is the band being used as the CMB map (see subsequent paragraph). $\Delta \Omega$ is the pixel size, $\sigma_{T_{ij}}^2$ is the average of $\sigma_{T_i}^2+\sigma_{T_j}^2$  over all pixels, $\bar{n}_{i_{tr}}$ is the average number of objects per steradian for the tracer, and $W_\ell^{T_i}$ includes the pixel and WMAP beam window functions.  $\mathbf{d}_X$ is the data vector for observable X with the entries in $\mathbf{d}_{\rm CO}$ and $\mathbf{d}_{Tr}$ being the CO fluctuation (in $\mu$K units) and the object number overdensity for the tracer, respectively. $\mathbf{S}_X$ is the fiducial signal covariance matrix and $\mathbf{N}_X$ is the noise covariance matrix with $\VEV{\mathbf{d}_X\mathbf{d}_X^T}=\mathbf{S}_X+\mathbf{N}_X$.  $(\mathbf{S}_X+\mathbf{N}_X)^{-1}$ works as an operator that weights each data point by its covariance, effectively filtering out very noisy modes.  It is evaluated using a multigrid preconditioner (see Appendix A of \citet{2007PhRvD..76d3510S} for details).  The variance for the estimator in Eq.~\ref{E:clest} is given by
\begin{eqnarray} \label{E:clestvar}
{\rm Var}[\hat{C}_\ell^{CO-Q}] = \frac{1}{(2\ell+1)f_{\rm sky}}\left[(\hat{C}_\ell^{CO-Q})^2+\hat{D}_\ell^{\rm CO}\hat{D}_\ell^Q\right]\, ,
\end{eqnarray}
where
\begin{eqnarray}
\hat{D}_\ell^{\rm CO}&=&\sum_{m=-\ell}^\ell \frac{|\bar{a}_{\ell m}^{\rm CO}|^2}{(2\ell+1)f_{\rm sky,CO}}\nonumber\\
\hat{D}_\ell^{Tr}&=&\sum_{m=-\ell}^\ell \frac{|\bar{a}_{\ell m}^{Tr}|^2}{(2\ell+1)f_{\rm sky,Tr}}\, ,
\end{eqnarray}
and $f_{\rm sky,CO}=0.783$, $f_{\rm sky,Q}=0.17$, and $f_{\rm sky,LRG}=0.203$ are the sky coverage fractions of our WMAP and SDSS data samples, respectively.  For completeness, we also give
\begin{eqnarray}\label{E:varcotr}
{\rm Var}[\hat{D}_\ell^{\rm CO}] = \frac{2}{(2\ell+1)f_{\rm sky,CO}}\left(\hat{D}_\ell^{\rm CO}\right)^2\nonumber\\
{\rm Var}[\hat{D}_\ell^{Tr}] = \frac{2}{(2\ell+1)f_{\rm sky,Tr}}\left(\hat{D}_\ell^{Tr}\right)^2\, .
\end{eqnarray}
By estimating the errors this way, we include all sources of noise including galactic foregrounds.

\begin{figure}
\begin{center}
{\scalebox{.3}{\includegraphics{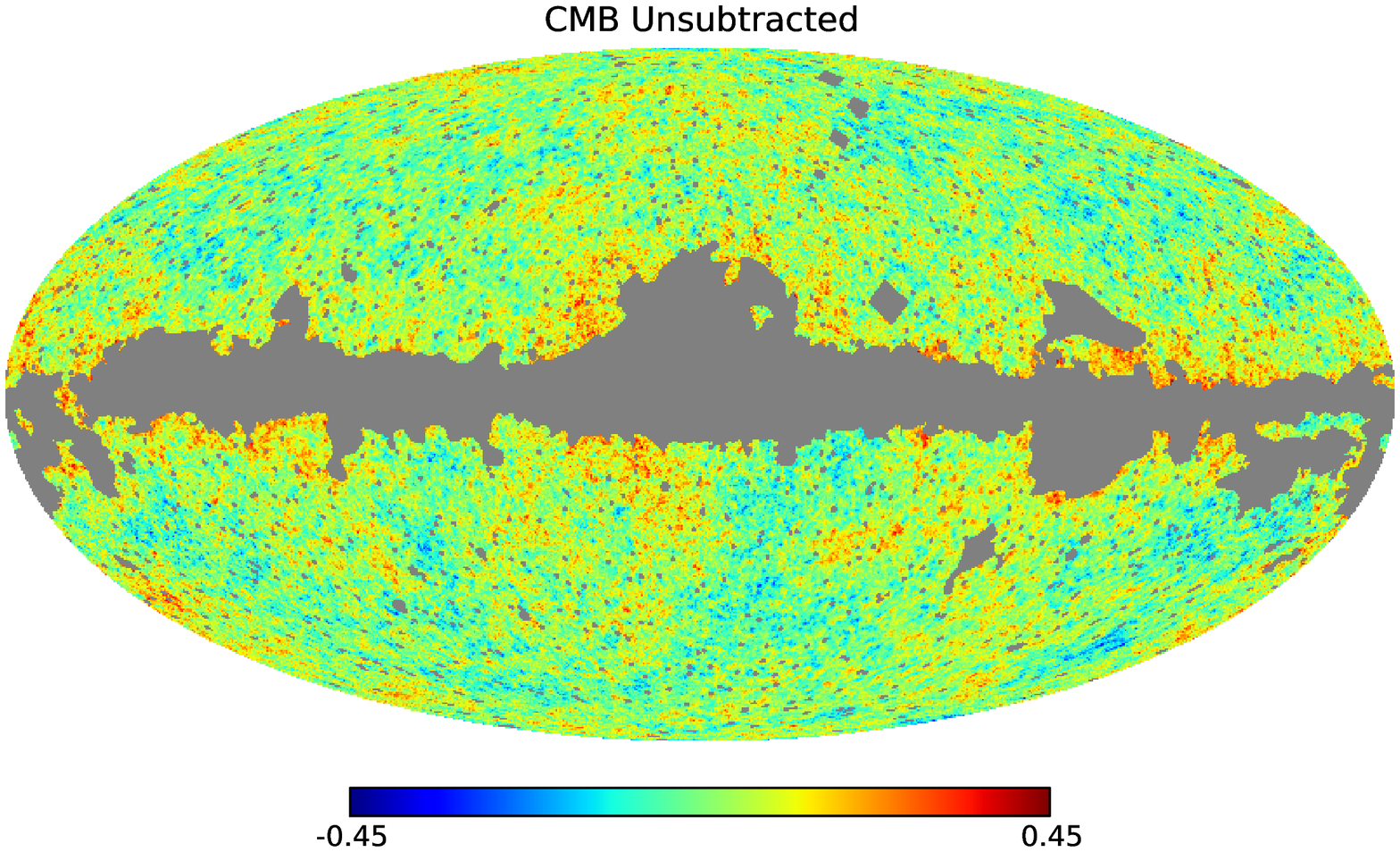}}}
{\scalebox{.3}{\includegraphics{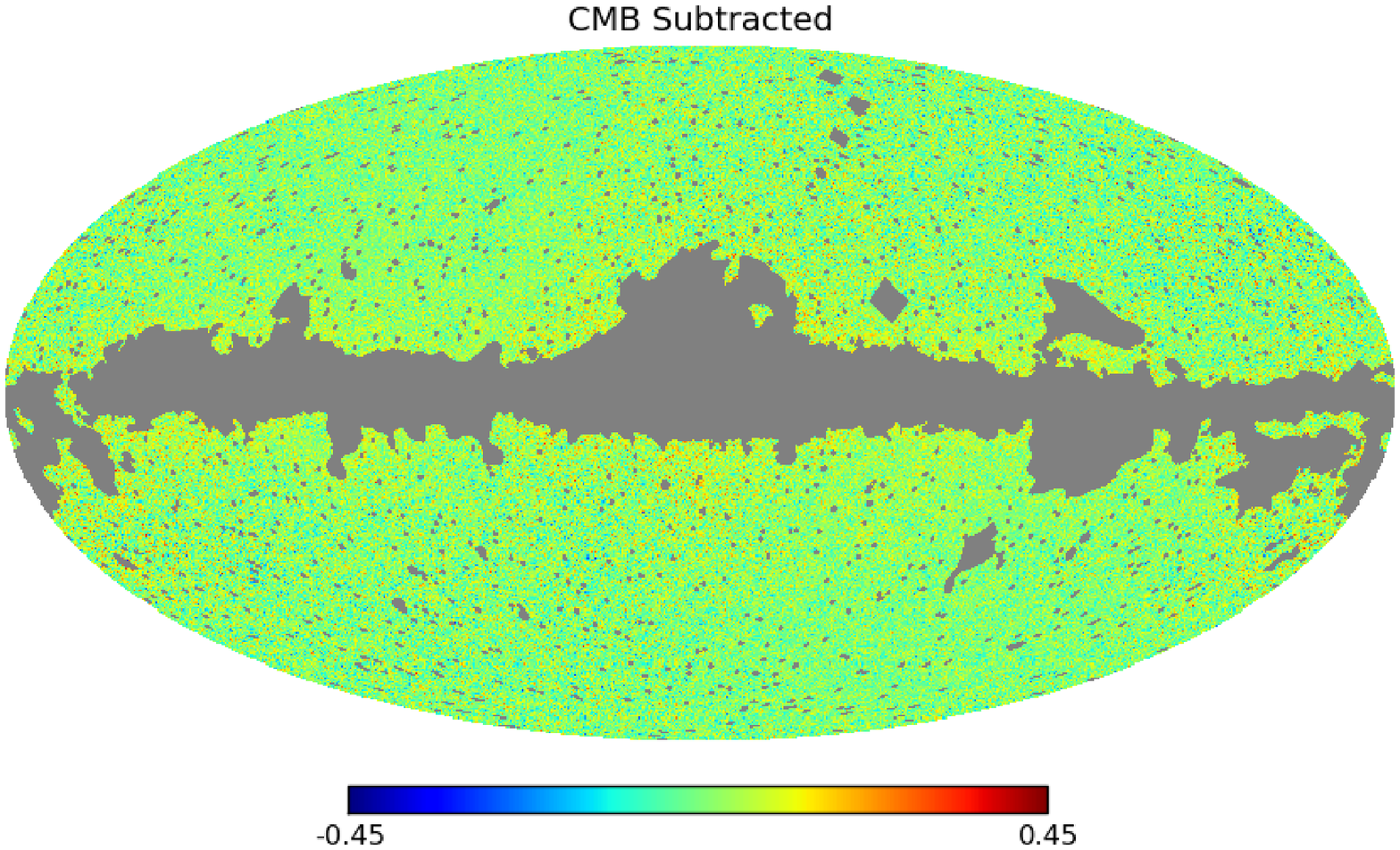}}}
\end{center}
\caption{\label{F:cocmb} The brightness temperature map in the WMAP Q band (left) and the same map with the brightness temperature map from the W band (estimating the CMB fluctuations) subtracted (right) with galactic emission and point sources masked out.  The color legend is given in units of mK.}
\label{fig:wmap_map}
\end{figure}

As discussed above, the WMAP temperature maps are dominated by the CMB and galactic foregrounds, with CO emission constituting a small contribution.  CMB fluctuations would drastically increase statistical errors in our CO search, so we choose to try and subtract out the CMB.  Because the WMAP bands are so large, we cannot model the CMB fluctuations to subtract them properly.  However, since the V and W maps are dominated by the CMB outside the galactic mask\footnote{Specifically, the foreground contribution to the V and W maps are approximately 20\%.}, we can use these maps to subtract the CMB from all the maps.  Specifically, we subtract the W map from the maps Ka and V, and we subtract the V map from the Q and W maps. Note that all the maps have zero mean.  Since the WMAP maps are already given in physical temperatures, we can just subtract them directly before converting them to brightness temperatures (see Sec.~\ref{S:wmaptemp}). The noise fluctuations in the V and W maps make the subtractions imperfect, even if we were to assume the foregrounds in these maps were negligible.  This causes noise fluctuations to increase due to the noise introduced by the subtracted map according to Eq.~\ref{E:clnco}, but this should not introduce a bias.  However, subtracting different bands, even with the extra noise, is worthwhile because keeping the CMB perturbations would contribute much more noise.  We give an example of a CMB-subtracted map in Fig.~\ref{F:cocmb}.  From the two maps, it is evident that our method removes the hot and cold spots of the CMB. Note that this subtraction does not affect the CO signal we aim at cross-correlating with since the bands do not overlap in CO redshift sensitivity as is clear in Tab.~\ref{T:zbin}.

\section{Results} \label{S:results}

\subsection{Quasar Results}
\begin{figure}
\begin{center}
\includegraphics[width=0.45\textwidth]{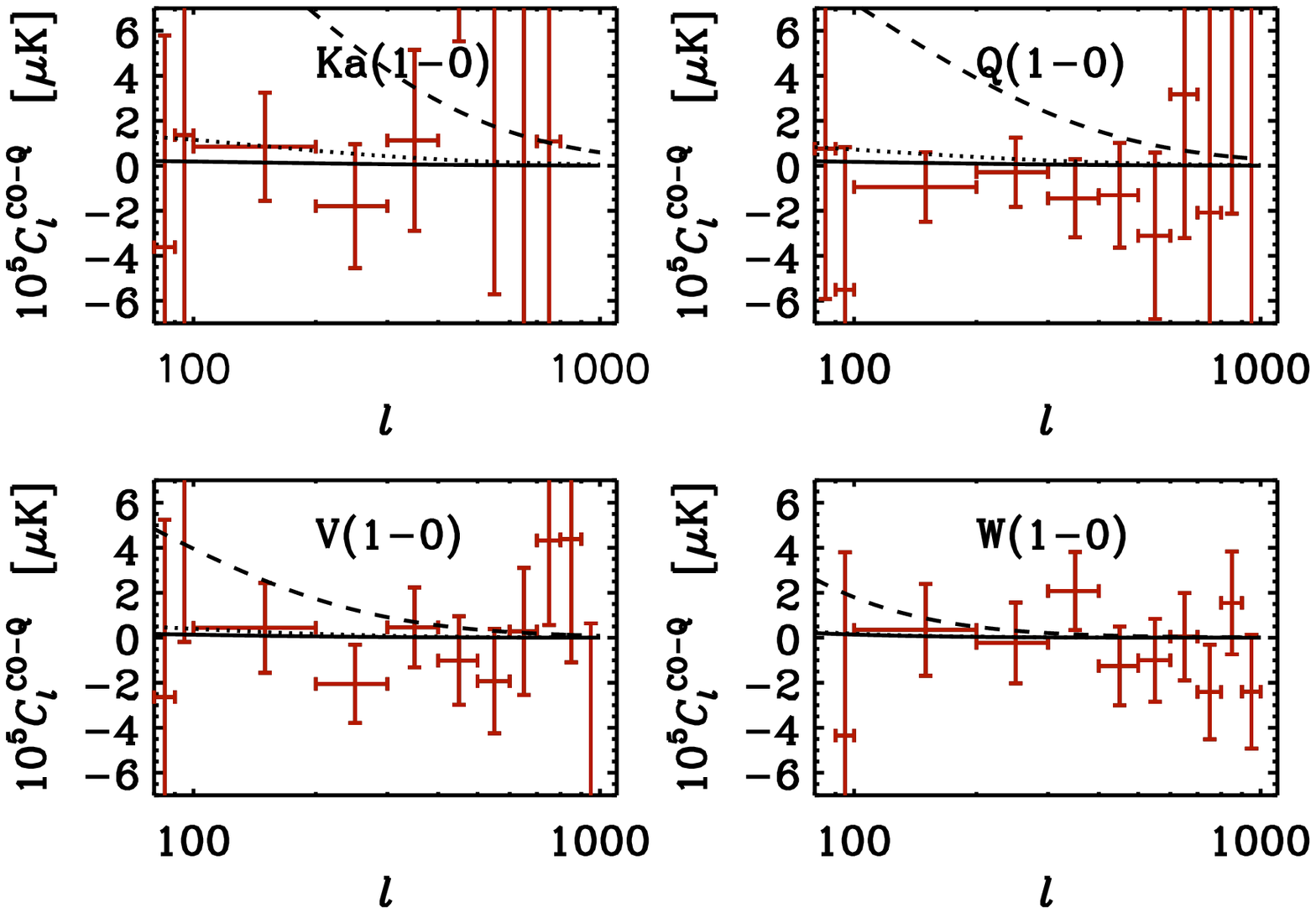}
\includegraphics[width=0.45\textwidth]{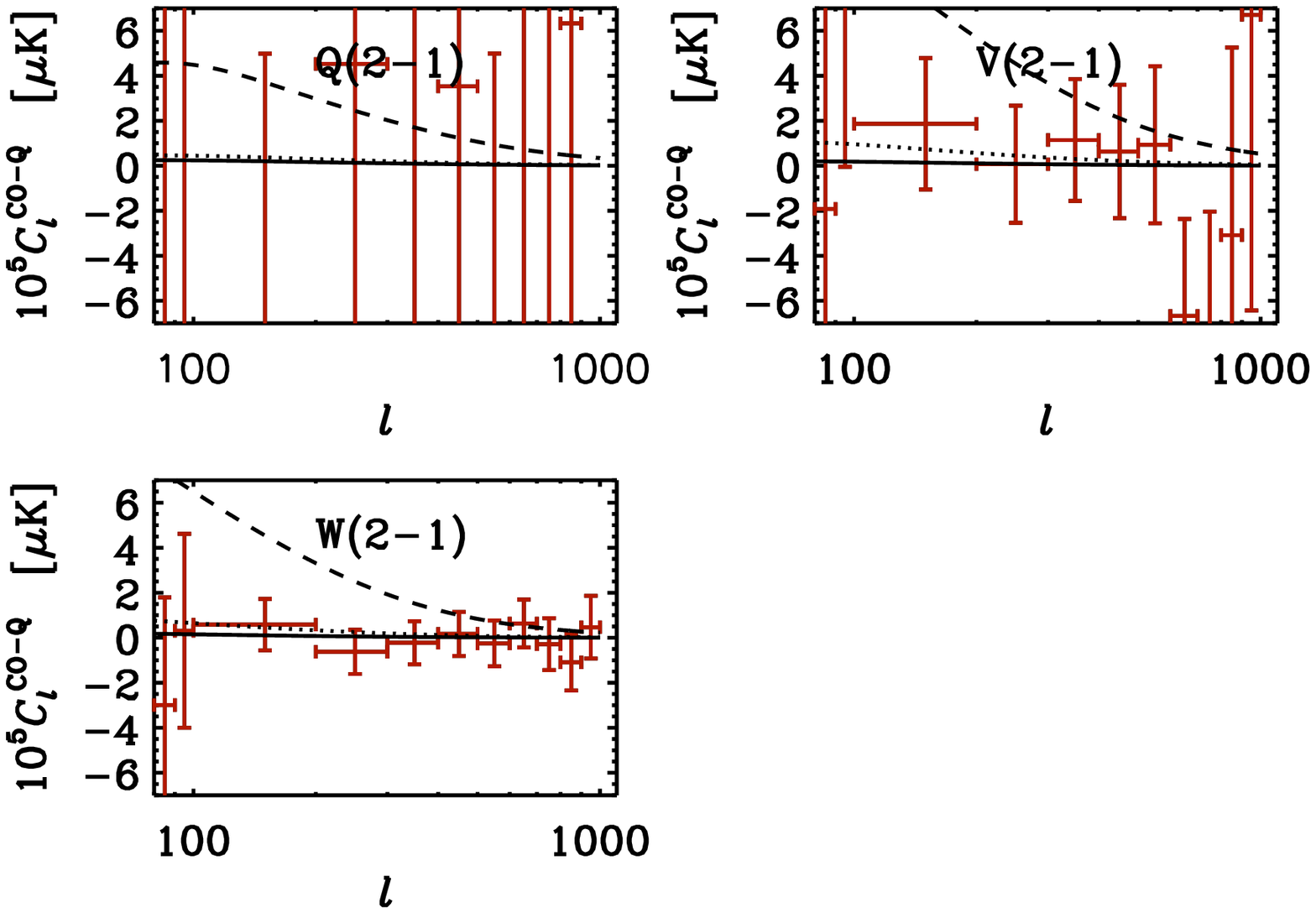}
\caption{\label{F:cross} The measured, binned CO-Q cross-correlation angular power spectrum measurements with 1-sigma error bars. The solid line is the nonzero theoretical CO-Q correlation from Model A for $M_{\rm co,min}=10^9M_\odot$, the dotted line is the rough estimate for Model B for $SFR_{\rm min}=0.01M_\odot/$yr, and the dashed line is 10x the estimate for Model B.}
\end{center}
\end{figure}

We present measurements of the CO-Q cross-correlation in Fig.~\ref{F:cross}.  Remember that the Ka band measurement for the CO(2-1) line was not performed.  Even though the CMB was subtracted from the data, the error bars on the cross-correlation are still much larger than the predicted spectra from Model A.  The WMAP temperature noise and resolution, as well as the number density of quasars and any remaining fluctuations due to foregrounds, affect the statistical errors, but from Fig.~\ref{F:cln} we can infer the noise from the brightness temperature residuals is by far the greatest contributor compared with its theoretical noise-free fluctuations.  It appears that our measured angular power spectra are consistent with a null signal and Model A in all the redshift bins.  Model B is fairly high in some of the bands, but not entirely ruled out.  We also show 10x the Model B estimate, which we see is ruled out in several of the bands on the larger scales.  This becomes much clearer in the $\VEV{T_{\rm CO}}$ constraints below.  Thus we can infer that the CO brightness is not significantly brighter than Model B ({\it e.g.}~there is no abundant population of faint CO emitters that is not included in our model, unless the cross-correlation coefficient $r_Q$ is low). However, a future experiment with an increased temperature sensitivity and quasar density is needed to detect the CO-Q cross-correlation (see Sec.~\ref{S:forecasts}).

\begin{figure}[!t]
\includegraphics[width=0.45\textwidth]{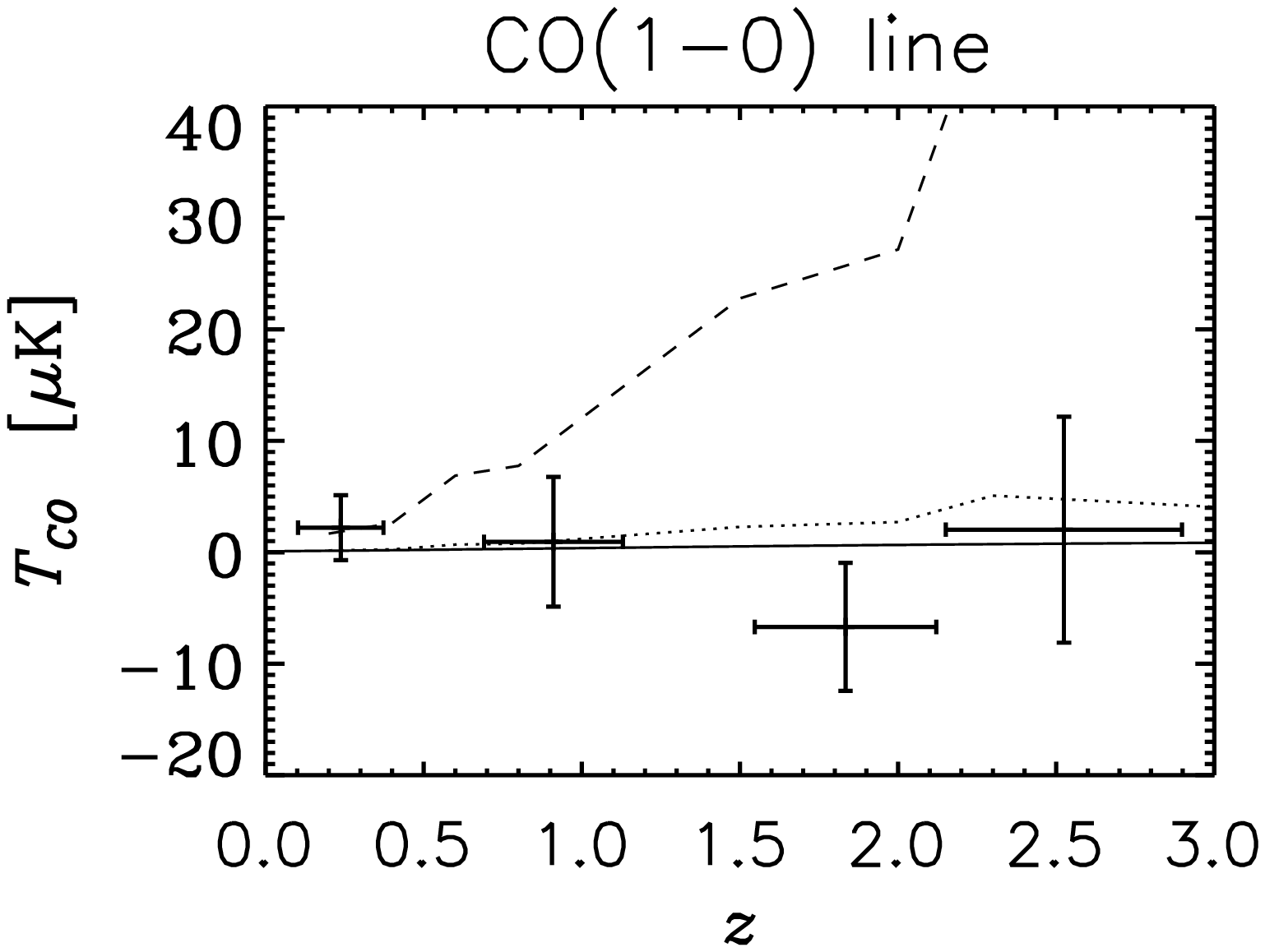}
\includegraphics[width=0.45\textwidth]{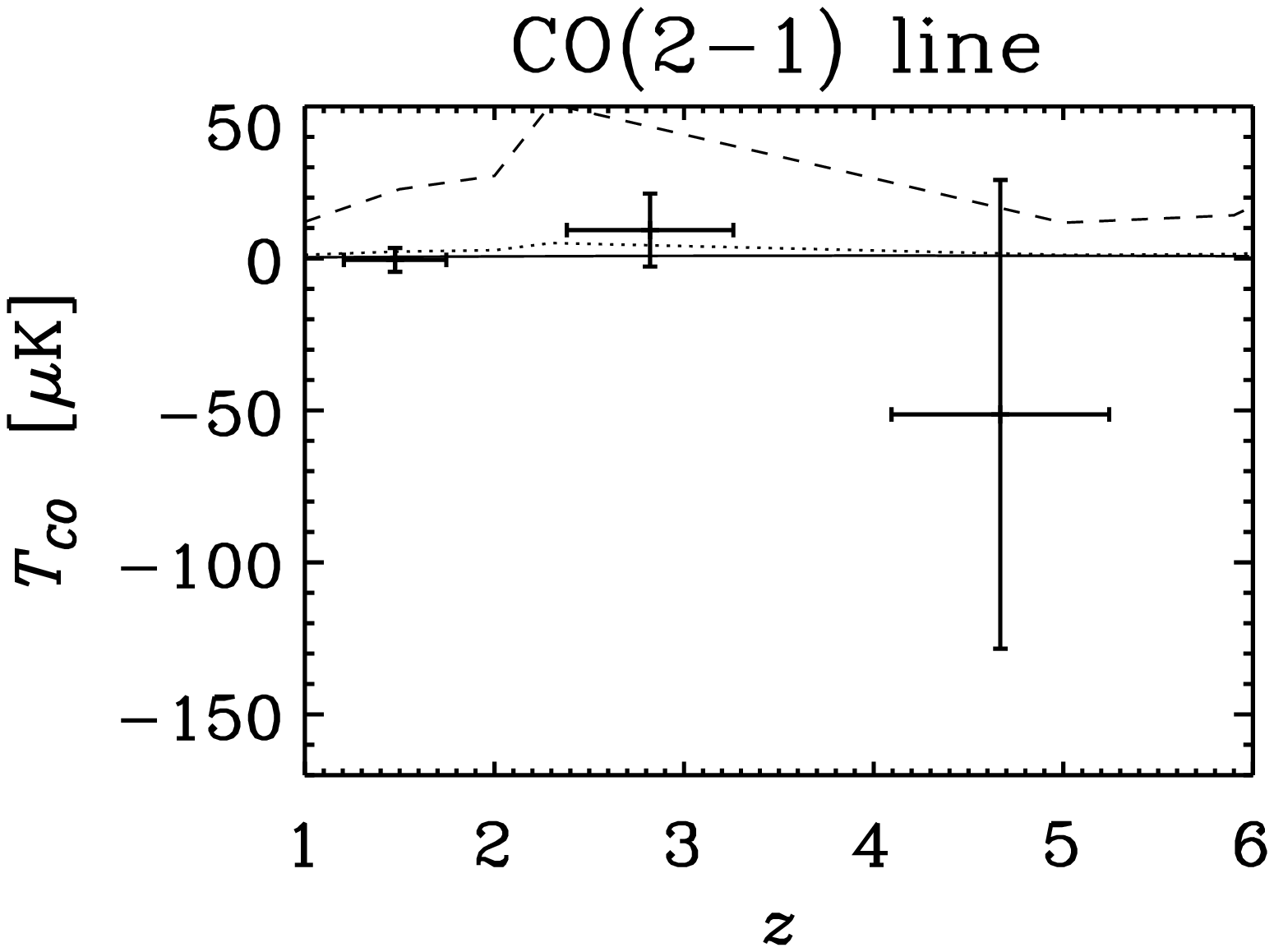}
\caption{\label{F:tcolim} Limits on $\VEV{T_{\rm CO}}$ of CO(1-0) and CO(2-1) over four redshift bins. The solid line is the fiducial temperature from Model A for $M_{\rm co,min}=10^9M_\odot$, the dotted line is the fiducial temperature from Model B for $SFR_{\rm min}=0.01M_\odot/$yr, and the dashed line is 10x the estimate for Model B.}
\end{figure}

Using the extra simplifying assumption that $\VEV{T_{\rm CO}}$ is constant over each redshift window we can translate our measurement into a constraint on the effective $\VEV{bT_{\rm CO}}$ temperature, $\overline{\VEV{bT_{\rm CO}}}$.  We define the $\VEV{bT_{\rm CO}}$ estimator by rewriting Eq.~\ref{Eq:ClCOQ}  as 
\begin{eqnarray}
\label{Eq:ClCOQ2}
C_\ell^{CO-Q}&=& r\overline{\VEV{bT_{\rm CO}}}\int dz \frac{H(z)}{c}\frac{f_{\rm CO}(z)f_Q(z)}{\chi^2(z)}b_Q(z)\times\nonumber\\
&&P_{\rm lin}(k=\ell/\chi(z),z)\nonumber\\
&=&r\overline{\VEV{bT_{\rm CO}}}F_\ell\, .
\end{eqnarray}
Assuming $r=1$, we then can construct a minimum-variance estimator for $\overline{\VEV{bT_{\rm CO}}}$ according to the expression
\begin{eqnarray} \label{E:btest}
\overline{\VEV{bT_{\rm CO}}}=\frac{\sum_\ell \hat{C}_\ell^{CO-Q}F_\ell/{\rm Var}[\hat{C}_\ell^{CO-Q}]}{\sum_\ell F_\ell^2/{\rm Var}[\hat{C}_\ell^{CO-Q}]}\, ,
\end{eqnarray}
with uncertainty
\begin{eqnarray} \label{E:btunc}
\frac{1}{\sigma_{bT_{\rm CO}}^2}=\sum_\ell\frac{F_\ell^2}{{\rm Var}[\hat{C}_\ell^{CO-Q}]}\, .
\end{eqnarray}
In practice, we limit ourselves to $80<\ell<1000$ to avoid the Integrated Sachs-Wolfe (ISW) ``contamination''  on larger scales \citep{1967ApJ...147...73S}.  Using the same formalism as above, our constraints on $\overline{\VEV{bT_{\rm CO}}}$ (assuming $r=1$) are given in Table \ref{T:btco}. If we further assume knowledge of the bias, which in practice means that we know that host halo mass of CO emitting objects, these constraints yield $\overline{T}_{\rm CO}$ constraints, where we rewrite Eq.~\ref{Eq:ClCOQ} as
\begin{eqnarray}
\label{Eq:ClCOQ3}
C_\ell^{CO-Q}= r\overline{T}_{\rm CO}\int dz \frac{H(z)}{c}\frac{f_{\rm CO}(z)f_Q(z)}{\chi^2(z)}b(z)b_Q(z)\times&&\nonumber\\P_{\rm lin}(k=\ell/\chi(z),z)\, .\,\,\,\,\,\,\,\,&&
\end{eqnarray}
Still assuming $r=1$, we plot constraints of $\VEV{T_{\rm CO}}$ for each line over four redshift bins, comparing the result to Model A for $M_{\rm co,min}=10^9M_\odot$, Model B for $SFR_{\rm min}=0.01M_\odot/$yr, in Fig.~\ref{F:tcolim}, and 10x Model B.  It is evident from these plots that the $\VEV{T_{\rm CO}}$ signal from Model A and Model B are not detectable from our analysis, but we know that the brightness cannot be an order of magnitude greater than Model B.  In fact, a model more than 3 times greater than Model B would be ruled out by the combined constraints of the two higher redshift points for the CO(1-0) line.  Another way of saying this is that the signal-to-noise ratio (SNR) for the two points, which is the SNRs of the two individual points added in quadrature, where the signal is the difference between the data point and the model, is more than 3 for a model more than 3 times Model B. 
\begin{table}[!t]
\begin{center}
\caption{\label{T:btco} Measured $\overline{bT}_{\rm CO}$ with 1-sigma error bars of CO emission lines $J=1\to0$ and $J=2\to1$ from cross-correlations of WMAP temperature bands with SDSS DR6 quasar (QSO) counts.}
\begin{tabular}{ccc}
\hline
WMAP band&$\overline{bT}_{CO(1-0)}\,(\mu$K)&$\overline{bT}_{CO(2-1)}\,(\mu$K)\\
\hline
Ka&$3.89\pm19.39$&N/A\\
Q&$-11.5\pm9.8$&$-138.\pm207.$\\
V&$1.41\pm8.68$&$18.7\pm24.1$\\
W&$2.98\pm3.94$&$-0.74\pm6.36$\\
\hline
\end{tabular}\end{center}
\end{table}

\subsection{LRG Results}\label{S:lrgresult}
Before measuring CO parameters using the LRG correlation, we proceed to determine the LRG clustering bias $b_{LRG}$ for the full sample, ignoring the bias redshift evolution. We start by rewriting the LRG angular auto-spectrum in Eq.~\ref{E:clauto} as
\begin{eqnarray}
C_\ell^{LRG}&=&b_{LRG}^2\int dz \frac{H(z)}{c}\frac{f_{LRG}^2(z)}{\chi^2(z)}P_{\rm lin}[k=\ell/\chi(z),z]\nonumber\\
&=&b_{LRG}^2G_\ell^{LRG}\, ,
\end{eqnarray}
where we assume a constant LRG bias.  We use a model for the measured LRG auto-spectrum given by $\hat{D}_\ell^{LRG}=b_{LRG}^2G_\ell^{LRG}+C_\ell^{LRG,{\rm shot}}$, where $C_\ell^{LRG,{\rm shot}}$ is the shot noise component for LRGs given in Eq.~\ref{E:shot}.  This model allows us to construct the estimator for the LRG bias
\begin{eqnarray}
[\widehat{b^2}]_{LRG}=\frac{\sum_\ell (\hat{D}_\ell^{LRG}-C_\ell^{LRG,{\rm shot}})G_\ell^{LRG}/{\rm Var}[\hat{D}_\ell^{LRG}]}{\sum_\ell (G_\ell^{LRG})^2/{\rm Var}[\hat{D}_\ell^{LRG}]}\, ,
\end{eqnarray}
with uncertainty
\begin{eqnarray}
\frac{1}{[\sigma_{\widehat{b^2}}]_{LRG}^2}=\sum_\ell\frac{(G_\ell^{LRG})^2}{{\rm Var}[\hat{D}_\ell^{LRG}]}\, ,
\end{eqnarray}
where we limit the sum to $10<\ell<100$.  Using this method, we find $b_{LRG}=2.482\pm0.055$, which is a little high compared to other analyses but still consistent given that our redshift range is wider.  In \citet{2008PhRvD..78d3519H}, the measured LRG bias for the relevant redshift range was $b_{LRG}=2.03\pm$0.07.  We list results assuming both values.

\begin{figure}
\begin{center}
\includegraphics[width=0.45\textwidth]{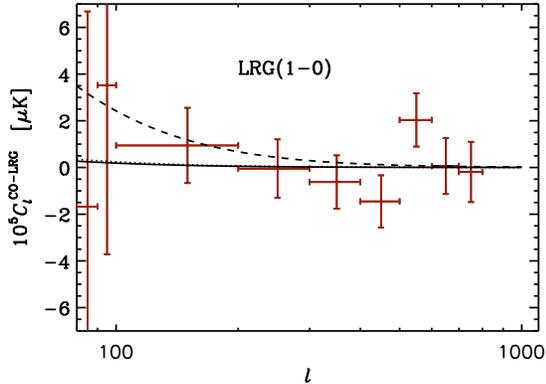}
\caption{\label{F:cldlrg} The measured, binned CO-LRG cross-correlation angular power spectrum measurements with error bars. The solid line is the nonzero theoretical CO-Q correlation from Model A for $M_{\rm co,min}=10^9M_\odot$, the dotted line is the rough estimate for Model B for $SFR_{\rm min}=0.01M_\odot/$yr, and the dashed line is 10x the estimate for Model B.}
\end{center}
\end{figure}

With an LRG bias, we can perform the CO-LRG cross-correlation measurement, the results of which we show in Fig.~\ref{F:cldlrg}.  These measurements, like the CO-quasar measurements, are consistent with both models, as well as 10x Model B, making it not really useful.  We estimate CO parameters for $0.16<z<0.47$ in a similar manner as the previous section.  The values we measured for $\overline{bT}_{CO(1-0)}$ and $\overline{T}_{CO(1-0)}$ assuming our LRG bias are $0.76\pm1.06$ $\mu$K and $0.56\pm0.78$ $\mu$K, respectively. Assuming the LRG bias from \citet{2008PhRvD..78d3519H}, the values change to $0.93\pm1.30$ $\mu$K and $0.69\pm0.96$ $\mu$K, respectively. Note that the cross-correlation measurement was performed without including the measured LRG bias to the fiducial signal; however, we repeated the measurement with the bias and confirmed that the final result was unchanged.  These constraints are indeed tighter than those from the quasars, due to a much higher areal density of objects.  However, at $z\simeq 0.25$, the relevant redshift, the predicted $\VEV{T_{\rm CO}}$ for both models are in the 0.1--0.2 range, making these constraints still not useful.

\subsection{Possible Contaminants} \label{S:results:SZ}

Our measurements could be contaminated by any extragalactic foregrounds correlated with quasars or LRGs.  We already attempted to remove ISW by neglecting large-scale modes, but other foregrounds may contaminate our signal.  For example, quasars could live in SZ-contributing clusters or be closely correlated with dust-rich galaxies, which would lead to a bias in our measurements.  Since we subtract maps, it is possible for a negative bias to occur, canceling a true signal.  Although this is unlikely, if nothing else it is useful to describe these possible contaminants since they will be important in future analyses.  We will briefly discuss these possibilities below.

As mentioned in Sec.~\ref{S:wmaptemp}, the thermal Sunyaev-Zel'dovich effect (tSZ) \citep{1980ARA&A..18..537S}  is expected to correlate CMB temperature with large scale structure tracers like quasars or LRGs.  In fact, there are hints of a SZ-quasar cross-correlation in WMAP/SDSS data \citep{2010ApJ...720..299C}.  Thus, it is important to check the level of tSZ contamination in our measurement.  Since tSZ has a well-defined frequency dependence, we can rule out tSZ if our cross-correlation measurements do not follow this frequency dependence.  Furthermore, we note that while performing the CMB subtraction using the V or W band, we also alter the tSZ signal frequency dependence so that at every frequency, $X$, we will write the tSZ amplitude, as $\Delta 
\tilde{T}_{\rm SZ} =\Delta T_{\rm SZ}^X-\Delta T_{\rm SZ}^{V\ or\ W}$  . Even though we do not expect tSZ in our CO analysis to be significant, in the following analysis we confirm whether or not this effect is indeed a problem.

The tSZ effect causes the CMB temperature to receive a secondary perturbation $\Delta T_{\rm SZ}$ every time it scatters with an object, with the perturbation given by
\begin{eqnarray}
\Delta T_{\rm SZ}(x) = yT_{\rm CMB}f(x)\, ,
\end{eqnarray}
where $y$ is the object-dependent Compton $y$ parameter, $x=h\nu/kT$, and $f(x)=x\coth x-4$.  This implies that once the CMB reaches us, its tSZ perturbation in a pixel due to a set of objects (of one type) in a redshift bin will be $\delta T=\Delta T_{\rm SZ}(N-\bar{N})$, where $N$ is the number of objects in the pixel and $\bar{N}$ is the average number of objects per pixel.  This model allows us to relate the angular cross-correlation between tSZ and a LSS tracer, $Tr$, to the tracer's auto-spectrum according to the form
\begin{eqnarray}
C_\ell^{SZ-Tr}=\Delta \tilde{T}_{\rm SZ}\bar{N}(C_\ell^{Tr}+C_\ell^{Tr,{\rm shot}})\, .
\end{eqnarray}
Using this model, we can construct an estimator for $\Delta \tilde{T}_{\rm SZ}$ using quasars as the LSS tracer, given by
\begin{eqnarray}
\hat{\Delta \tilde{T}_{\rm SZ}}=\frac{1}{\bar{N}}\frac{\sum_\ell \hat{C}_\ell^{SZ-Q}(C_\ell^{Q}+C_\ell^{Q,{\rm shot}})/{\rm Var}[\hat{C}_\ell^{SZ-Q}]}{\sum_\ell (C_\ell^{Q}+C_\ell^{Q,{\rm shot}})^2/{\rm Var}[\hat{C}_\ell^{SZ-Q}]}\, .
\end{eqnarray}
The procedure for measuring the tSZ-Q cross-spectrum is equivalent to measuring the CO-Q cross-spectrum, except that tSZ-Q uses the CMB physical temperature while CO-Q uses the brightness temperature.  This allows us to set $\hat{C}_\ell^{SZ-Q}=\hat{C}_\ell^{CO-Q}/f_{\rm br}$, where $f_{\rm br}$ is the prefactor in Eq.~\ref{E:brcon} for converting physical temperatures to brightness temperatures.

\begin{figure}[!t]
\includegraphics[width=0.45\textwidth]{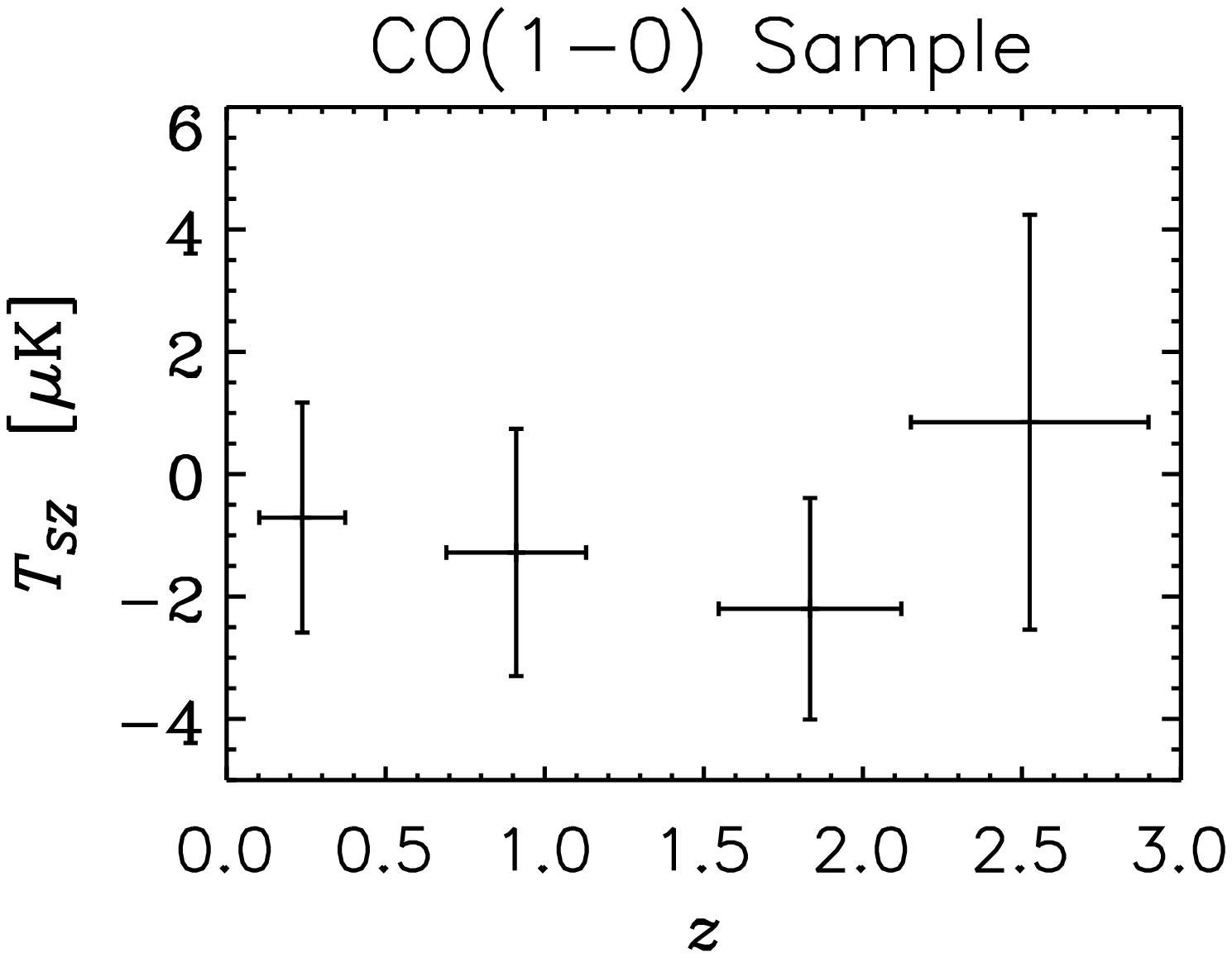}
\includegraphics[width=0.45\textwidth]{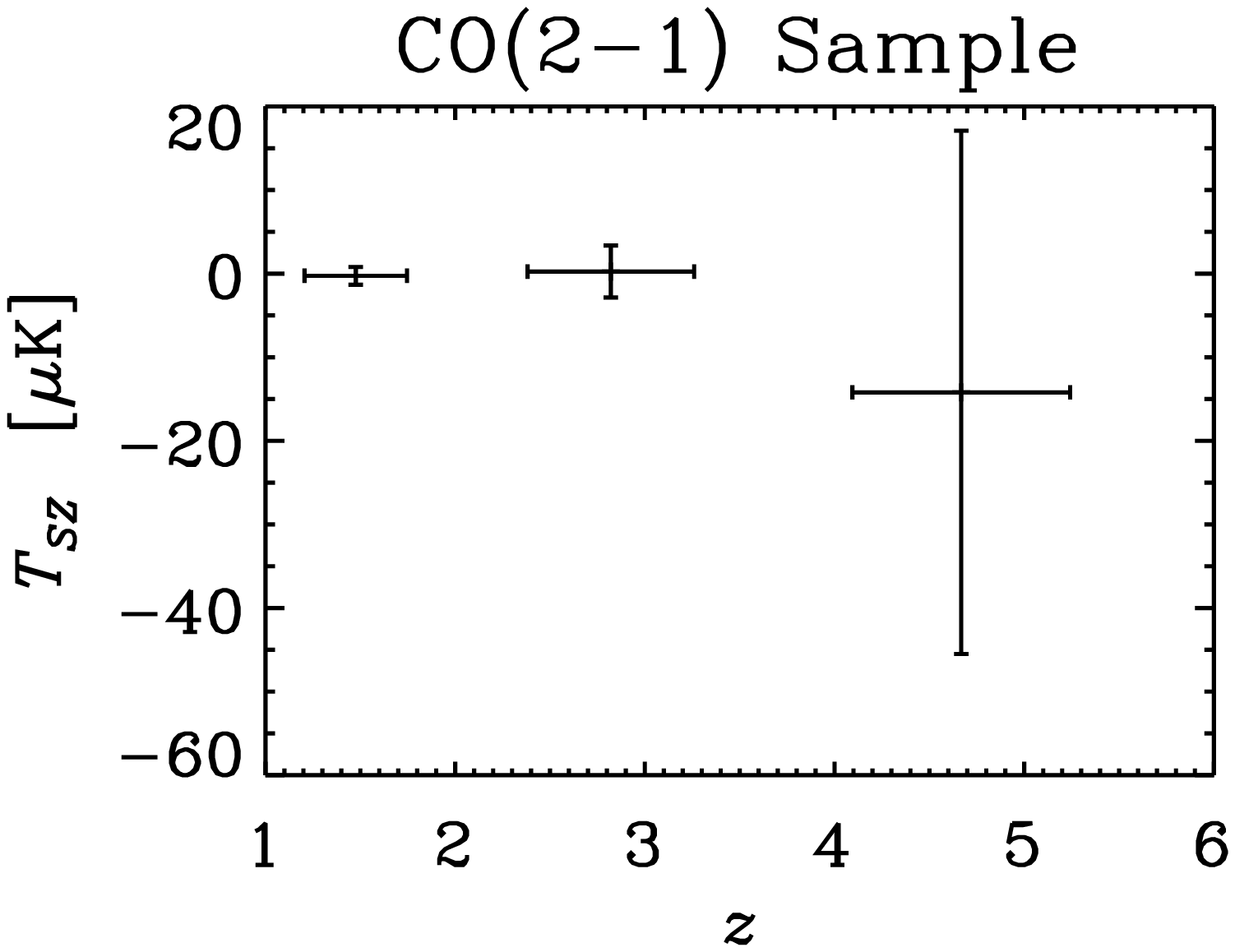}
\caption{\label{F:tsz} Limits on $\Delta \tilde{T}_{\rm SZ}$ for the CO(1-0) and CO(2-1) quasar samples. For the CO(1-0) samples, the points from left to right to right are for the WMAP maps W-V, V-W, Q-W, and Ka-W.  For the CO(2-1) samples, the points from left to right to right are for the WMAP maps W-V, V-W, and Q-W.  The $z$ axis labels the redshift of the quasar map.}
\end{figure}

We show the measured $\Delta \tilde T_{\rm SZ}$ at each redshift for the quasar samples in Fig.~\ref{F:tsz}.  For the LRG sample, we find $\Delta \tilde T_{\rm SZ}=\{0.54\pm1.05,0.61\pm1.12\}$ for $b_{LRG}=\{2.48,2.03\}$, respectively.  None of these measurements are significant, but since we did not detect CO emission, we can't rule out the presence of tSZ either.  We emphasize that it is necessary to search for tSZ if CO line emission is detected in any future experiment.  We note that a spectrograph like the one we propose in Sec.~\ref{S:forecasts} will have a high enough resolution to easily remove tSZ (and the CMB) directly, so this should not be an issue.

\begin{figure}
\begin{center}
\includegraphics[width=0.45\textwidth]{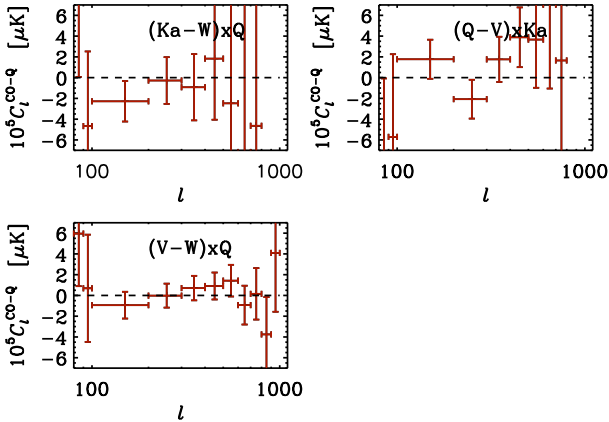}
\caption{\label{F:cldust} The measured, binned CO-Q cross-correlation angular power spectrum measurements for out-of-band quasars with 1-sigma error bars.  This measurement shows that the temperature correlation with dust is not significant enough to bias the CO-emission measurements. The dashed line represents a null correlation.}
\end{center}
\end{figure}

We also investigated contamination by dusty galaxies. Dust has a positive spectral index, causing its contribution with the W and V bands to be higher than the other bands.  Thus, while correlating the difference of two WMAP bands, e.g. Q-V, with quasars, we may get a negative correlation which could possibly cancel the CO-quasar correlation. We test this possibility by correlating Ka-W, Q-V, and V-W with quasars that would correlate with the CO(1-0) line in the Q, Ka, and Q bands, respectively.  Since this correlation cannot come from CO, we know that if dusty galaxies are canceling the CO-quasar correlation, they would create a negative signal.  Note that W-V correlated with quasars or LRGs should 
not have this problem because the signal would be positive.  We find the correlations are indeed consistent with zero, as seen in Fig.~\ref{F:cldust}, so we can conclude that dusty galaxies are not canceling a CO-quasar cross-correlation detection. Similarly, this null result also rules out other possible contaminants such as radio sources.

\subsection{Another use of WMAP data}

Another method of detecting CO temperature fluctuations is by cross-correlating the CO(1-0) and CO(2-1) lines coming from the same emission redshift.  For each WMAP band, the CO(1-0) redshift range and the CO(2-1) redshift range are different; this can cause the CO(1-0) redshift range for one band to overlap with the CO(2-1) redshift range of another band, e.g. Ka for CO(1-0) and V for CO(2-1).  Since these two rotational lines would presumably be emitted by the same sources, we can cross-correlate pairs of WMAP bands that overlap in this way to search for CO emission. This could be a way to measure the line ratio as a function of redshift. Specifically, we could search for an angular cross-correlation of the form
\begin{eqnarray}
C_\ell^{CO-CO'}=\int dz \frac{H(z)}{c}\frac{f_{\rm CO}(z)f_{CO'}(z)}{\chi^2(z)}\times&&\nonumber\\sP_{\rm CO}[k=\ell/\chi(z),z]\, ,\,\,\,\,\,\,\,\,&&
\end{eqnarray}
integrated over the intersection of the two redshift ranges where unprimed CO is CO(1-0) and primed CO is CO(2-1).  We include a parameter $s$, the ratio between the two CO lines.  We can attempt to measure a unitless amplitude $B$ such that $\hat{C}_\ell^{CO-CO'}=BC_\ell^{CO-CO'}$.  In this case, $B$ has an estimator similar to Eqs.~\ref{E:btest} and \ref{E:btunc}.  In this procedure, there would be other sources of correlation in the WMAP bands coming from various foregrounds, with the CMB as the dominant foreground.  To remove this complication, we can either remove these foregrounds by hand or cross-correlate pairs of WMAP bands that do not correspond to the associated CO redshift window as a cross-check to see if the difference in cross-correlation between the two cases is statistically significant. A third option would be to use a three point function, something like Ka-V-QSO, as this would lead to an extra handle on the foreground correlation.

\begin{figure}
\begin{center}
\includegraphics[width=0.45\textwidth]{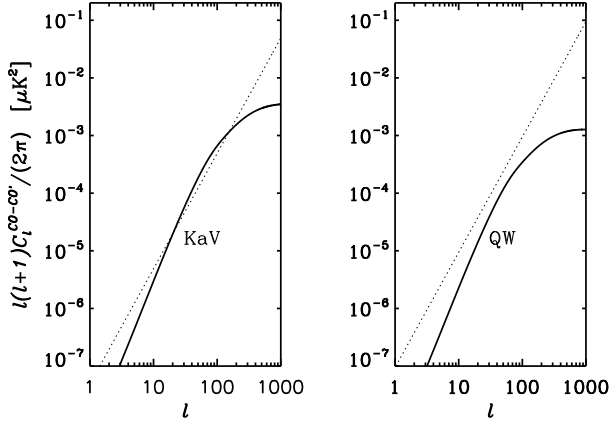}
\caption{\label{F:cl12} The predicted angular cross-power spectra from Model A for $M_{\rm co,min}=10^9M_\odot$ with a WMAP band for CO(1-0) cross-correlated with another WMAP band for CO(2-1).  The solid line is the clustering term and the dotted line is the shot noise term.}\end{center}
\end{figure}

Based on the WMAP band CO redshift ranges in Table \ref{T:zbin}, we find that the Ka band for CO(1-0) should overlap with the V band for CO(2-1), while the Q band for CO(1-0) should overlap with the W band for CO(2-1).  The other combinations (Ka-Q, Ka-W, Q-V, and V-W) should not exhibit any cross-correlations that are due to common CO emission.  The predicted spectra for Ka-V and Q-W are plotted in Fig.~\ref{F:cl12}.  We attempt to subtract the CMB from all four bands, cross-correlate the Ka-V and Q-W pairs, and then look for a detection of an amplitude.  However, when we calculate the variance of the cross-correlation $C_\ell^{CO-CO'}$ given by
\begin{eqnarray}\label{E:clttestvar}
{\rm Var}[\hat{C}_\ell^{CO-CO'}] = \frac{1}{(2\ell+1)f_{\rm sky,CO}}\left[(\hat{C}_\ell^{CO-CO'})^2+\right.&&\nonumber\\
\left.\hat{D}_\ell^{\rm CO}\hat{D}_\ell^{CO'}\right]\, ,\,\,\,\,\,\,\,\,&&
\end{eqnarray}
we find extremely large errors for this CO amplitude.  
The errors for the amplitude for Model A are $\sigma_B\simeq1400$ for Ka-V and $\sigma_B\simeq 1200$ for Q-W.  The numbers are much better for Model B, being $\sigma_B\simeq45$ for Ka-V and $\sigma_B\simeq 68$ for Q-W, although still less powerful than those from the WMAP-quasar cross-correlation.
The limits from the COxQSO analysis greatly outperform the limits from a potential WMAP CO(1-0)xCO(2-1) analysis because the instrumental noise is  relatively much higher than the shot noise as shown in Fig.~\ref{F:cln}.   In addition, WMAP foregrounds and CO emission from higher $J$ can contaminate the signal, so we do not attempt to measure it in this analysis.

\section{CO Experiment Forecasts} \label{S:forecasts}
Now that we have determined WMAP's limited ability to constrain CO temperature, we investigate what can be done in future experiments.  We know that the limiting factor in the CO-Q cross-correlation measurement is how well we can reduce temperature noise as well as increase the quasar density.  Unlike WMAP, a dedicated experiment should have small frequency bands in order to maximize power from the CO line and to subtract continuum sources, including the CMB and other foregrounds.  We consider such a configuration here\footnote{Mike Seiffert, private communication}.

In this section, we seek to determine the signal-to-noise ratio (SNR) of the cross-correlation between the brightness temperature fluctuation of CO(1-0) measured by a hypothetical CO spectrograph and the latest spectroscopic LSS surveys at $z\sim3$.  This CO spectrograph measures the CO(1-0) rotational line at observed frequency 28.8 GHz with a 1 GHz bandwidth and 20 50-MHz frequency channels.  At a frequency resolution of approximately $R=600$, this experiment should be able to effectively model CMB and foreground emission and remove their contributions.  The central frequency of the band would correspond to $z=3$ with channel widths corresponding to $\Delta z = 0.00694$ ($2.9306<z<3.0694$).  Table \ref{T:copar} shows sample parameters of this experiment for two cases.  For each case, we set a instrumental error $\sigma_{\rm T,fwhm}$, a beam FWHM, and a survey area.  Note that $\sigma_{\rm T,fwhm}$ is for a pixel corresponding to the FWHM we choose. We then have a noise angular power spectra of the form
\begin{eqnarray}
C_\ell^{n,CO}=\sigma_{\rm T,fwhm}^2(0.4245\theta_{\rm fwhm})^2/W_\ell^T\, .
\end{eqnarray}
\begin{table}
\begin{center}
\caption{\label{T:copar} CO spectrograph parameters.}
\begin{tabular}{ccc}
\hline
Survey Parameter&Spec 1&Spec 2\\
\hline
$\sigma_{\rm T,fwhm}$ $(\mu$K)&6.7&4.7\\
$\theta_{\rm fwhm}$ (arcmin)&40.5&28.3\\
Survey Area (deg$^2$)&550&270\\
\hline
\end{tabular}\end{center}
\end{table}

We begin with the BOSS final spectroscopic quasar survey.  The BOSS full survey will cover 10,200 deg$^2$ by 2014, long before the time when the CO spectrograph would start observing.  In the range compatible with our hypothetical spectrograph, based on projections from the BOSS Year One (partial) survey, we expect BOSS to detect 5465 spectro-QSOs per steradian.  Based on the redshift distribution in \citet{2012ApJS..199....3R}, we see that the distribution is pretty flat in our range of interest, making us set the density in each redshift bin equal to 5465/20 = 273 str$^{-1}$.

In principle, the spectral resolution is such that the cross-correlation will be performed directly in three dimensions, as is currently performed with 21cm surveys \citep{2010Natur.466..463C,2013ApJ...763L..20M} and L11. For the sake of simplicity, we will however compute here a simple forecast using a bin-by-bin 2D angular power spectra.  Using the 2D angular power spectrum to constrain CO temperature neglects modes along the line of sight, lowering the SNR.  Thus, we take the  SNR forecasts in this section to be lower limits to what could be achieved with a full 3D power spectrum analysis.

The SNR for the experiment is then
\begin{eqnarray}
{\rm SNR}^2 =N_{\rm z-bins}\sum_\ell\frac{(C_\ell^{CO-Q})^2}{{\rm Var}[C_\ell^{CO-Q}]}\, ,
\end{eqnarray}
where $N_{\rm z-bins} = 20$ and we assume each bin contributes approximately equally.  Note that we neglect the cross-correlation shot noise in this calculation, as well as for the rest of this Section.  We find for Model A with $M_{\rm co,min}=10^9M_\odot$ the forecasts SNR = 1.2 (5.4) for Spectrograph 1 and SNR = 1.8 (8.0) for Spectrograph 2 for each redshift bin (over the full redshift range).  Spectrograph 2 may be of interest for the full redshift range, but systematics resulting from not subtracting SZ and dust contamination properly may degrade the signal.  An autocorrelation would have an even lower SNR (SNR $\sim7.1$ for Spectrograph 2).  Probing Model B should have better prospects; although we can't estimate the cross-spectrum directly, we can estimate that since $\VEV{T_{\rm CO}}$ at $z=3$ for Model B with $SFR_{\rm min}=0.01M_\odot/$yr is about $4/0.85\simeq4.8$ higher than for Model A with $M_{\rm co,min}=10^9M_\odot$, its cross-spectrum will be approximately 4.8 times higher.  This changes the SNR values to 4.2 (18.8) for Spectrograph 1 and  4.4 (19.7) for Spectrograph 2.  Note that the SNR does not increase by a factor of 4.8 because the noise is model-dependent.  The autocorrelation SNR would actually be higher than that for the cross-spectra in this case (SNR $\sim49$ for Spectrograph 2), but foregrounds can degrade this signal.  Another option would be to have an instrument that probed CO fluctuations at two frequencies with one twice the frequency of the other.  With this, you could search for a cross-correlation signal from CO(1-0)$\times$CO(2-1); the SNR for an ``equivalent Spectrograph 2" for this setup is approximately 59 without the foreground issues of an autocorrelation.  However, this would be a much more expensive instrument.

Instead of using BOSS, we could also cross-correlate Spectrographs 1 and 2 with HETDEX \citep{2008ASPC..399..115H}, which will observe 1 million Ly$\alpha$ emitters over 200 square degrees in the redshift range $1.8<z<3.8$.  In a $\Delta z=0.007$ redshift bin, HETDEX will have an areal density of about 57,500 str$^{-1}$.  This specification cross-correlated with Spectrograph 2 gives us an SNR for each bin (over full redshift range $2.93<z<3.07$) of 5 (22) for Model A and 13 (58) for Model B, which is much higher than we can get with BOSS.  Using instead Spectrograph 1 decreases the numbers only slightly, to 3.8 (17) for Model A and 13 (57) for Model B.

Another option is to just cross-correlate the temperature maps from the Planck satellite \citep{2011A&A...536A...1P} with the same photo-quasar map we used in this analysis.  However, we find that for Model A none of bands have a SNR greater than one.  We find that the highest SNR we get is 0.73 for the CO(2-1) line for Planck's 143 GHz band.  We assume subtracting the 100 GHz band from all the other bands and subtracting the 70 GHz band from the 100 GHz band.  All the other SNRs for the other Planck bands are much less for the both lines.  Model B is more constrainable with SNR=2.0 in the 143 GHz band for the CO(2-1) line, but foregrounds will probably degrade this signal. Using the BOSS spectro-quasar survey would increase $f_{\rm sky}$ to at most 0.25.  Since most of the noise errors is due to the temperature maps, Planck x BOSS would not do much better than Planck x SDSS DR6.  Moreover, the use of the CII line instead of CO would be more appropriate for Planck high frequencies.

Current ground-based, high-angular-resolution CMB polarimeters offer another promising avenue towards measuring this contribution. Consider an SPTPol-like survey \citep{2012SPIE.8452E..1EA} with 6.5 and 4.5 $\mu$K-arcmin sensitivity at 90 and 150 GHz with  2.0 and 1.2 arcmin FWHM and covering 500 sq.~degrees of a BOSS-like quasar survey.  The real SPTPol and BOSS surveys cover opposite hemispheres, but we perform this exercise as an illustration.   We assume both bands have a 30\% bandwidth.  For $z\sim3$ quasars, the 90 and 150 GHz bands could constrain $\VEV{T_{\rm CO}}$ for $J=$3 and 5, respectively.  The BOSS full survey will have quasar areal densities within the relevant redshift bins of about 36,000 and 22,000 str$^{-1}$, respectively.  With these parameters, we forecast cross-correlation SNRs for Model A (Model B) of 5 (12) and 8 (13).  For the extended SPT-3G, the 220 GHz band is added, which can constrain the $J=8$ line using quasars at $z\sim3$.  The survey area is also increased to 2500 sq.~degrees.  The sensitivities for the 90, 150 and 220 GHz bands change to 4.2, 2.5, and 4.0 $\mu$K-arcmin and the FWHMs change to 1.7, 1.2, and 1.0 arcmin.  In the 220 GHz band, the areal density of quasars in the relevant redshift bin is about 15,000 str$^{-1}$.  For SPT-3G, we forecast cross-correlation SNRs for Model A (Model B) of 15 (31), 24 (31) and 54 (59).  HETDEX does not do as well as BOSS in this case because it covers such a small area.  Of course, SPTPol cannot constrain the same lines as the 28.8 GHz experiment mentioned earlier.  Also, the large bands and limiting frequency coverage will make the foreground removal difficult so that these numbers are optimistic. Also, the high optical depth limit is less certain for higher $J$ lines, making the signal more uncertain for this experiment.  Finally, the frequency bands for ACPol/SPTPol are wide enough that a full 3D analysis is not feasible.

We summarize the various results for different experiment combinations in Table \ref{T:snr}.  Other experiments that could also help with this kind of search are the Primordial Inflation Explorer (PIXIE) \citep{2011JCAP...07..025K}, and the Murchison Widefield Array (MWA) \citep{2009IEEEP..97.1497L}.  Although PIXIE is a polarization experiment to detect inflationary gravitational waves, its high frequency resolution over a wide frequency range can constrain CII and CO, particularly for higher $J$ lines, over large redshift ranges, including high redshifts.  MWA is currently searching for the 21 cm HI line from the dark ages and reionization.  As mentioned in L11, CO x 21 cm can be a powerful probe of the high redshift universe.  For example, MWA x SPTPol could constrain CO J=5 or 7 lines and star formation at $z\sim6-7$.
\begin{table*}
\begin{center}
\caption{\label{T:snr} The signal-to-noise ratio (SNR) for measuring the CO brightness temperature with CO$\times$LSS cross-correlations of various experiment combinations.  Unless noted otherwise, the first and second value listed for the SNR are for Model A and Model B, respectively.}
\begin{tabular}{cccc}
\hline
&CO line&SNR per $\Delta z=0.007$&SNR over full $z$-range\\
\hline
Spectrograph 1 $\times$ BOSS QSOs&CO(1-0)&1.2, 4.2&5.4, 19\\
Spectrograph 2 $\times$ BOSS QSOs&CO(1-0)&1.8, 4.4&8.0, 20\\
Spectrograph 1 $\times$ HETDEX Ly$\alpha$ emitters&CO(1-0)&3.8, 13&17, 57\\
Spectrograph 2 $\times$ HETDEX Ly$\alpha$ emitters&CO(1-0)&5, 13&17, 58\\
\hline
Planck (143 GHz) $\times$ SDSS DR6 QSOs&CO(2-1)&N/A&2 (Model B)\\
\hline
SPT (90 GHz) $\times$ BOSS QSOs&CO(3-2)&N/A&5,12\\
SPT (150 GHz) $\times$ BOSS QSOs&CO(5-4)&N/A&8,13\\
SPT-3G (90 GHz) $\times$ BOSS QSOs&CO(3-2)&N/A&15,31\\
SPT-3G (150 GHz) $\times$ BOSS QSOs&CO(5-4)&N/A&24,31\\
SPT-3G (220 GHz) $\times$ BOSS QSOs&CO(8-7)&N/A&54,59\\
\hline
\end{tabular}\end{center}
\end{table*}

\section{Conclusions} \label{S:conclude}

We have predicted an angular cross-power spectrum between CO line emission and quasars and LRGs based on $\Lambda$CDM cosmology and the L11 model.  We proposed searching for the quasar/LRGs cross-correlation to characterize CO emission in high-redshift galaxies.  We have also attempted to detect the cross-correlation in WMAP and SDSS photo-quasars and LRGs up to $z\sim6$.  A signal was not detectable, mainly due to the large statistical errors in the WMAP maps.  We were able to set upper limits to the brightness temperature of the CO(1-0) and CO(2-1) lines, which rule out models much greater than our Model B.  We also explored the CO(1-0)xCO(2-1) cross correlation, another signature of CO emission.  
Although current probes appear to be unable to detect CO emission, the potential for future experiments looks considerably greater. Current or soon-to-happen ground based, high-angular-resolution CMB experiments overlapping  with BOSS offer a chance to detect higher $J$ lines.  In our forecasts for an optimistic model for a future spectrograph to detect CO(1-0) line emission, we found a SNR of 58 for a CO(1-0)x(HETDEX Ly$\alpha$ emitter) analysis at $z\sim3$ and a SNR of 59 for a more expensive CO(1-0)xCO(2-1) analysis.  Although these numbers will likely be decreased due to foreground subtraction, this result is still very promising.  A future detection of CO brightness temperature perturbations will allow us to model the CO emission-line galaxies at high redshifts, possibly even out to redshifts in the reionization epoch.

\begin{acknowledgments}

We thank D.~Hanson, S.~Furlanetto, and M.~Seiffert for helpful comments and useful discussions. Part of the research described in this paper was carried out at the Jet Propulsion Laboratory, California Institute of Technology, under a contract with the National Aeronautics and Space Administration. AP was supported by an appointment to the NASA Postdoctoral Program at the Jet Propulsion Laboratory, administered by Oak Ridge Associated Universities through a contract with NASA. This work was supported by the Keck Institute of Space Studies and we thank colleagues at the ``First Billion Years'' for stimulating discussions, in particular J. Bowman and A. Readhead for organizing it.

Funding for the SDSS and SDSS-II has been provided by the Alfred P. Sloan Foundation, the Participating Institutions, the National Science Foundation, the U.S. Department of Energy, the National Aeronautics and Space Administration, the Japanese Monbukagakusho, the Max Planck Society, and the Higher Education Funding Council for England. The SDSS Web Site is {\tt http://www.sdss.org/}. The SDSS is managed by the Astrophysical Research Consortium for the Participating Institutions. The Participating Institutions are the American Museum of Natural History, Astrophysical Institute Potsdam, University of Basel, University of Cambridge, Case Western Reserve University, University of Chicago, Drexel University, Fermilab, the Institute for Advanced Study, the Japan Participation Group, Johns Hopkins University, the Joint Institute for Nuclear Astrophysics, the Kavli Institute for Particle Astrophysics and Cosmology, the Korean Scientist Group, the Chinese Academy of Sciences (LAMOST), Los Alamos National Laboratory, the Max-Planck-Institute for Astronomy (MPIA), the Max-Planck-Institute for Astrophysics (MPA), New Mexico State University, Ohio State University, University of Pittsburgh, University of Portsmouth, Princeton University, the United States Naval Observatory, and the University of Washington.
\end{acknowledgments}


\end{document}